\documentclass[superscriptaddress,showpacs,preprintnumbers,amsmath,amssymb,twocolumn,prb]{revtex4-1}
\usepackage{graphicx}
\usepackage{dcolumn}
\usepackage{bm}
\usepackage{color}
\renewcommand{\bm}{\mathbf}

\begin{document}

\title{Third harmonic generation on exciton-polaritons in bulk semiconductors \\ subject to a magnetic field}

\author{W.~Warkentin}
\affiliation{Experimentelle Physik 2, Technische Universit\"{a}t Dortmund, D-44221 Dortmund, Germany}
\author{J.~Mund}
\affiliation{Experimentelle Physik 2, Technische Universit\"{a}t Dortmund, D-44221 Dortmund, Germany}
\author{D.~R.~Yakovlev}
\affiliation{Experimentelle Physik 2, Technische Universit\"{a}t Dortmund, D-44221 Dortmund, Germany}
\affiliation{Ioffe Institute, Russian Academy of Sciences, 194021 St. Petersburg, Russia}
\author{V.~V.~Pavlov}
\affiliation{Ioffe Institute, Russian Academy of Sciences, 194021 St. Petersburg, Russia}
\author{R.~V.~Pisarev}
\affiliation{Ioffe Institute, Russian Academy of Sciences, 194021 St. Petersburg, Russia}
\author{A.~V.~Rodina}
\affiliation{Ioffe Institute, Russian Academy of Sciences, 194021 St. Petersburg, Russia}
\author{M.~A.~Semina}
\affiliation{Ioffe Institute, Russian Academy of Sciences, 194021 St. Petersburg, Russia}
\author{M.~M.~Glazov}
\affiliation{Ioffe Institute, Russian Academy of Sciences, 194021 St. Petersburg, Russia}
\author{E.~L.~Ivchenko}
\affiliation{Ioffe Institute, Russian Academy of Sciences, 194021 St. Petersburg, Russia}
\author{M.~Bayer}
\affiliation{Experimentelle Physik 2, Technische Universit\"{a}t Dortmund, D-44221 Dortmund, Germany}
\affiliation{Ioffe Institute, Russian Academy of Sciences, 194021 St. Petersburg, Russia}

\date{\today}

\begin{abstract}
We report on a comprehensive experimental and theoretical study of optical third harmonic generation (THG) on the exciton-polariton resonances in the zinc-blende semiconductors GaAs, CdTe, and ZnSe subject to an external magnetic field, representing a topic that had remained unexplored so far. In these crystals, crystallographic THG is allowed in the electric-dipole approximation, so that no strong magnetic-field-induced changes of the THG are expected. Therefore, it comes as a total surprise that we observe a drastic enhancement of the THG intensity by a factor of fifty for the $1s$-exciton-polariton in GaAs in magnetic fields up to 10 T. In contrast, the corresponding enhancement is moderate for CdTe and almost neglectful for ZnSe. In order to explain this strong variation, we develop a microscopic theory accounting for the optical harmonics generation on exciton-polaritons and analyze the THG mechanisms induced by the magnetic field. The calculations show that the increase of THG intensity is dominated by the magnetic field enhancement of the exciton oscillator strength which is particularly strong for GaAs in the studied range of field strengths. The much weaker increase of THG intensity in CdTe and ZnSe is explained by the considerably larger exciton binding energies, leading to a weaker modification of their oscillator strengths by the magnetic field.
\end{abstract}

\maketitle

\section{Introduction}

In the broad field of fundamental and applied nonlinear optics, the
coherent processes of frequency conversion such as sum and difference frequency generation, are important tools for studying light-matter interaction and measuring nonlinear optical susceptibilities \cite{Shen,Boyd}. They have played a crucial role as effective tools for extending the emission of coherent light sources to longer or shorter wavelengths in the infrared or the ultraviolet range thus greatly expanding the availability of such sources for important applications.

The second and third optical harmonic generation (SHG and THG), as the lowest order nonlinear frequency conversion processes, are generally characterized by relatively high values of nonlinear susceptibilities in crystals. They have been widely investigated and used for developing versatile devices in nonlinear photonics. The SHG process is allowed in the electric-dipole approximation for noncentrosymmetric crystals. In case of fulfilling the phase-matching condition the transformation of the incident light with frequency $\omega$ into SHG light with frequency $2\omega$ can be highly efficient. However, in crystals with inversion symmetry SHG is forbidden. The higher order nonlinear process of THG at frequency $3\omega$
is much weaker, with typical susceptibilities of the order $|\chi^{(3)}|\sim 10^{-24}$ m$^2/$V$^2$ as compared to $|\chi^{(2)}|\sim10^{-12}$ m$/$V for SHG~\cite{Boyd}. However, the THG process is allowed in solids of any symmetry, as well as in amorphous materials, glasses, liquids, gases, and plasmas. Obviously, this universality of THG
does not release it from the restriction to fulfill the conservation laws of energy,
wave-vector and angular momentum in the elementary photon conversion.

Semiconductors have been always in the focus of optical harmonic generation studies. However, the investigations performed in the 1960-90ies were mostly limited to single wavelengths or to narrow spectral ranges.  Only recently with the availability of efficient optical parametric oscillators, spectroscopic studies of harmonics generation became more versatile. In particular, spectroscopic investigations with high spectral resolution at low temperatures and with application of electric and magnetic fields opened new experimental opportunities for obtaining important information on the role of excitons in the processes of nonlinear light-matter interaction and the relevant basis microscopic mechanisms  \cite{Pisarev, Yakovlev18}. Most of these studies were devoted to SHG in prototype semiconductors  GaAs \cite{Pavlov,Saenger06,Brunne15}, CdTe \cite{Saenger06}, ZnO \cite{Lafrentz13}, (Cd,Mn)Te \cite{Saenger06b}, EuTe, and EuSe \cite{Kaminski}. Recently, they were extended to two-dimensional semiconductors WSe$_2$ and MoS$_2$ \cite{Wang2015,Trolle2015,Glazov2017}. In contrast, up to now spectroscopic studies of THG in semiconductors have remained scarce, limited to reports for the magnetic semiconductors EuTe and EuSe\cite{Lafrentz10,Lafrentz12} and the electric field effect on exciton THG in  GaAs\cite{Brunne15}. The involvement of the third-order nonlinearity in the THG process allows one to study electronic states which are hidden in lower-order processes \cite{Akhmanov}. Since the THG process is allowed in all media in the electric-dipole approximation, which is the strongest mechanism of light-matter interaction, one might not expect a strong influence of moderate external fields on the THG signals.

In this paper, we present a spectroscopic study of THG involving the exciton-polariton states in GaAs, CdTe and ZnSe semiconductors in an external magnetic field. These materials have the same noncentrosymmetric zinc-blende crystal structure with $T_d$ as crystallographic point group (P.~G.), but their exciton binding energy increases from 4.2 to 10 and 20~meV going from GaAs to CdTe and further to ZnSe. We find, quite unexpectedly, a strong increase of the THG intensity by a factor of 50 in GaAs in magnetic fields up to 10~T. However, a much weaker effects are observed in CdTe and ZnSe. A theoretical model for THG on exciton-polaritons and its modification by a magnetic field is developed. It quantitatively describes the bulk of experimental data and shows that the increase of the exciton oscillator strength is the main factor in the THG intensity increase by the magnetic field.

\section{Experimental details}
\label{Experimental details}

\begin{figure}[t]
\centering
\includegraphics[width=0.42\textwidth]{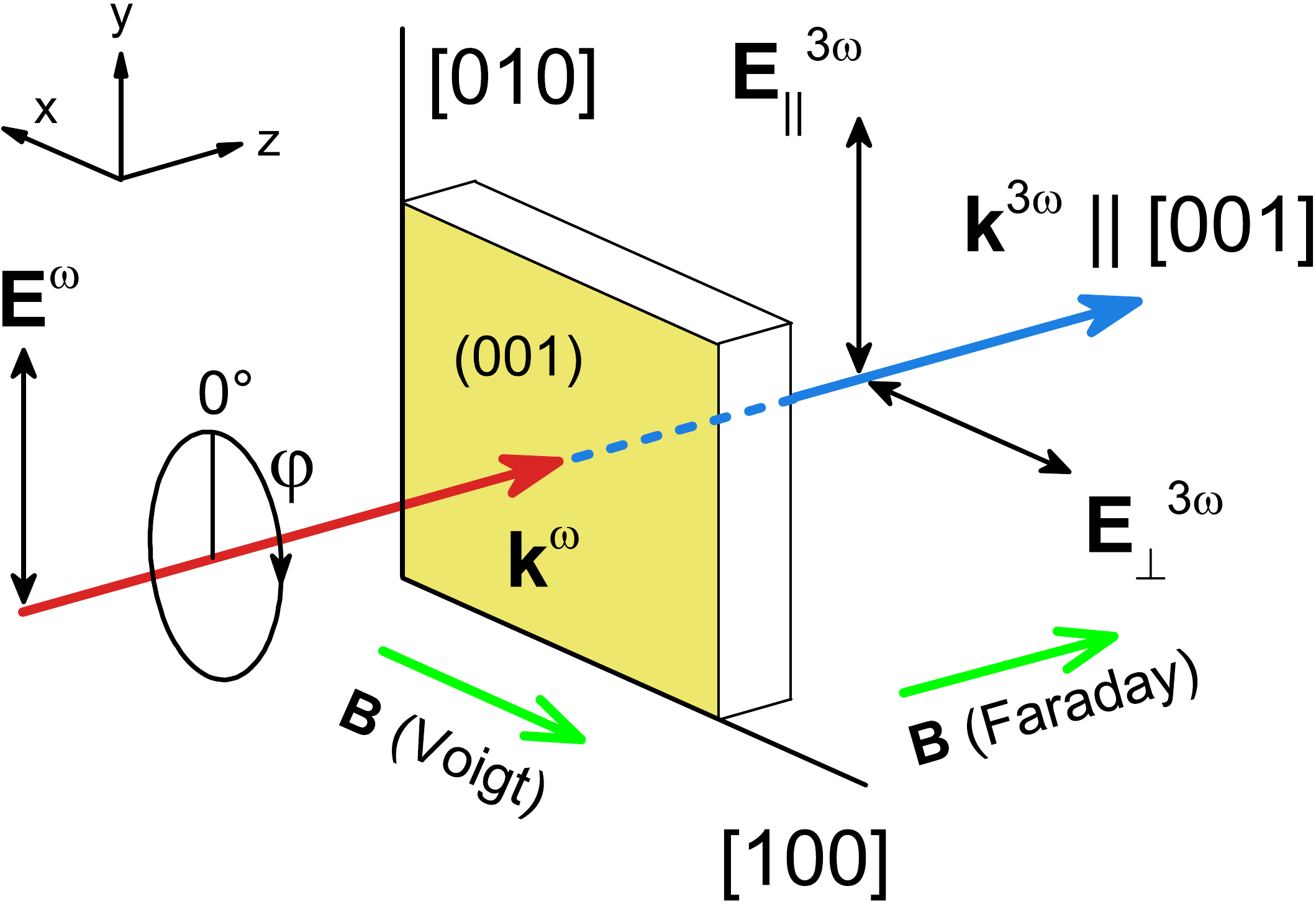}
\caption{Sketch of the experimental geometry in the THG studies of (001)-oriented GaAs, CdTe and ZnSe samples.}
\label{fig:Fig-1}
\end{figure}

THG spectra were recorded in transmission geometry with
$\mathbf{k}^\omega \parallel {z}$, where ${z}$ is the structure growth axis $[001]$. A sketch of the experimental geometry is shown in Fig~\ref{fig:Fig-1}. We used laser pulses of $7$~ns duration at $10$~Hz repetition rate, generated by an optical parametric oscillator pumped by the third harmonic of a solid-state Nd:YAG laser\cite{Saenger06}. The laser photon energy $\hbar\omega$ was tuned across a finite spectral range at about $E_g/3$, where $E_g$ is the semiconductor band-gap energy. The energy per pulse was set to $0.3$~mJ and the diameter of the focusing spot on the sample to $0.5$~mm.
The THG signals were spectrally separated from the fundamental light by bandpass optical filters and a monochromator and detected by a cooled charge-coupled-device (CCD) camera. External magnetic fields up to 10~T were applied in two geometries: the Voigt geometry $\mathbf{B \perp k}^\omega$, $\mathbf{B} \parallel [100]$ and the Faraday geometry $\mathbf{B \parallel k}^\omega \parallel [001]$. The incoming light was linearly polarized and the THG signal was measured either in parallel ($\mathbf{E}^{3\omega} \parallel \mathbf{E}^{\omega}$) or perpendicular ($\mathbf{E}^{3\omega} \perp \mathbf{E}^{\omega}$)
linear polarizations relative to the incoming light polarization. For measuring rotational anisotropy diagrams, the light polarization was controlled by half-wave plates oriented at an azimuthal angle $\varphi$. For $\varphi=0^\circ$, the light is polarized along the $y$ axis and for $\varphi=90^\circ$, along the $x$-axis, see Fig.~\ref{fig:Fig-1}.

The most detailed studies were performed on a GaAs sample. The 10-$\mu$m-thick GaAs layer was grown by the gas phase epitaxy on a semi-insulating GaAs (001) substrate\cite{Zhilyaev}. Magnetic-field-induced SHG in GaAs was reported in Refs.~\onlinecite{Pavlov,Saenger06} and the electric field effect on SHG and THG was shown in Ref.~\onlinecite{Brunne15}. Experiments were also performed on a CdTe epilayer and a ZnSe bulk sample. The 1-$\mu$m-thick CdTe layer was grown by molecular-beam epitaxy on a GaAs(001) substrate with a Cd$_{0.8}$Mg$_{0.2}$Te buffer layer \cite{Saenger06}. The CdTe layer was overgrown by a 50 nm Cd$_{0.8}$Mg$_{0.2}$Te cap layer to reduce undesired surface effects. The ZnSe sample, an (001)-oriented slab with a thickness of 2 mm, is a bulk crystal grown by the Bridgman method.

The linear optical properties of exciton resonances were assessed via reflectivity spectra measured with a halogen lamp. A back reflection geometry with nearly normal light incidence was chosen for that. In the Voigt geometry, the reflectivity was measured for linear polarizations parallel or perpendicular to the magnetic field direction, and in the Faraday geometry the circularly polarized reflection was analyzed. All measurements were performed for samples being in contact with helium gas at a temperature $T=5$~K.

\section{Experimental results}
\label{Experiment}

\subsection{Reflectivity spectra at the $1s$-exciton in GaAs in magnetic field}
\label{Reflectivity in GaAs}

\begin{figure}[t]
\centering
\includegraphics[width=0.45\textwidth]{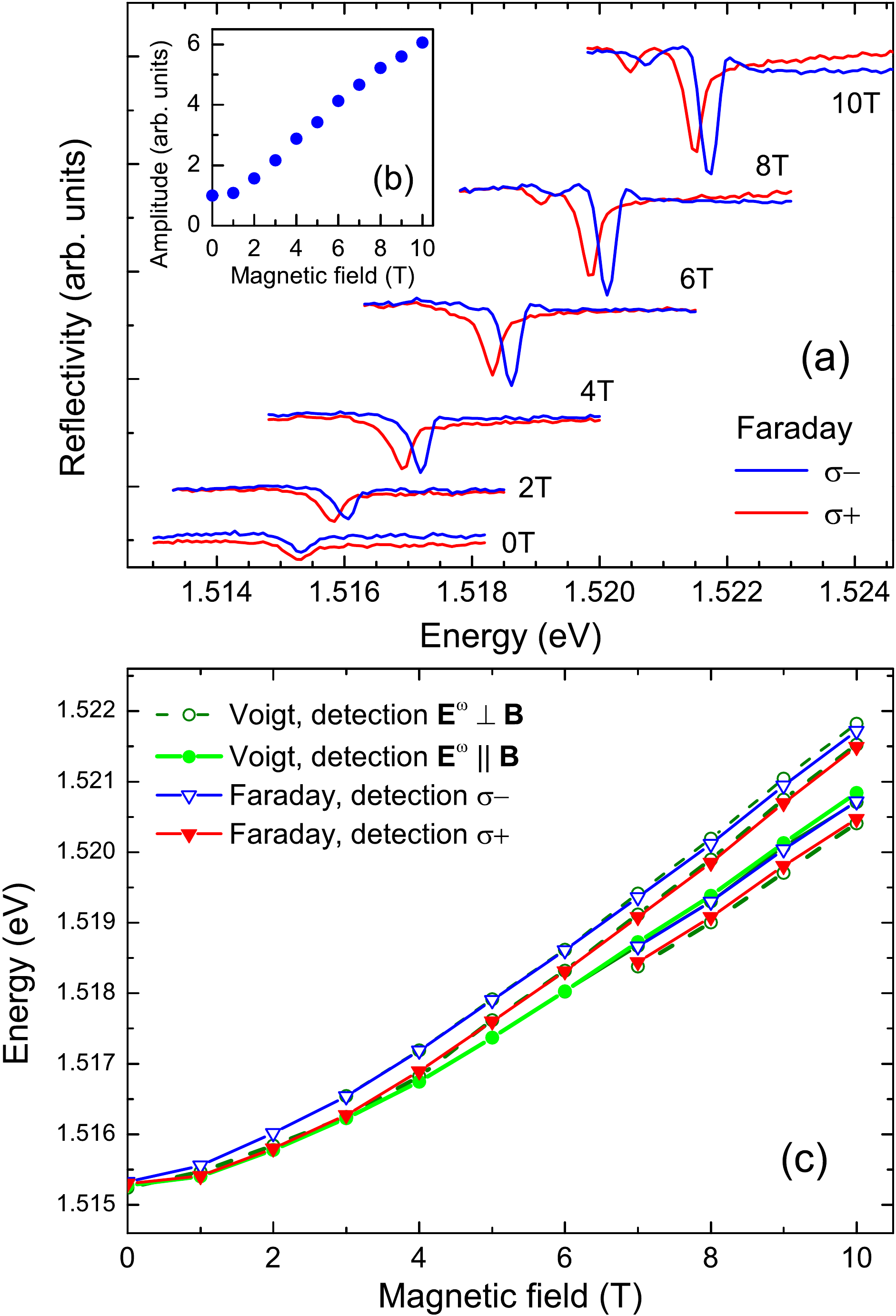}
\caption{(a) Reflectivity spectra in the vicinity of the $1s$-exciton in GaAs measured as a function of applied magnetic fields at $T=5$~K. Spectra are shifted vertically for clarity. (b) Magnetic field dependence of the amplitude of the reflectivity resonance normalized to its value at $B=0$, which reflects the growth of the exciton oscillator strength with $B$. (c) Diamagnetic energy shifts of exciton-polariton components in the Faraday and Voigt geometries. Symbols are experimental data and lines are interpolations.
}
\label{fig:Fig-2}
\end{figure}

The linear optical properties of GaAs exciton-polaritons in external magnetic fields are well studied experimentally and theoretically \cite{Altarelli1973,Dingle1973,Fischbach1974,Lipari1975,Swierkowski1975,Nam1976,Seisyan,Seisyan2016}. In order to have reference data for our THG results, we present in Fig.~\ref{fig:Fig-2} the reflectivity spectra measured in a magnetic field applied in the Faraday geometry. The spectrum has a dip at the energy of the $1s$-exciton. More precisely, this dip corresponds to the upper exciton-polariton branch, while the lower branch is not resolved due to the rather weak polariton effect in GaAs. The position of the upper polariton branch is just slightly above the $1s$-exciton. The evolution of the dip energy and its depth in external magnetic field is  directly related to the key exciton parameters. Here, we are interested in energy position, diamagnetic shift and oscillator strength of the exciton resonances in magnetic field.

The studied GaAs layer has high structural and optical quality\cite{Zhilyaev}, which is confirmed by the narrow exciton lines in the reflectivity spectra with widths of 0.3~meV, see Figs.~\ref{fig:Fig-2}(a) and \ref{fig:Fig-3}. With increasing magnetic field up to 10~T the exciton resonance shifts to higher energy and splits. The line shifts in the Faraday and Voigt geometries detected for various polarizations are given in Fig.~\ref{fig:Fig-2}(c). The identification and discussion of these lines, including the selection rules, can be found  in Refs.~\onlinecite{Altarelli1973,Dingle1973,Lipari1975} on the basis of the complex valence band structure resulting in geometry-dependent diamagnetic shifts and Zeeman effects. Figure~\ref{fig:Fig-2}(b) shows the magnetic field dependence of the amplitude of the reflectivity signal, where the dip depth is normalized to its value at $B=0$ for the $1s$-exciton resonance. The oscillator strength increases by a factor of 6 with the field growing from $0$ to 10~T. We will use this parameter for modeling the THG signals.

\subsection{Third harmonic generation in GaAs}
\label{THG in GaAs}

\begin{figure}[t]
\centering
\includegraphics[width=0.48\textwidth]{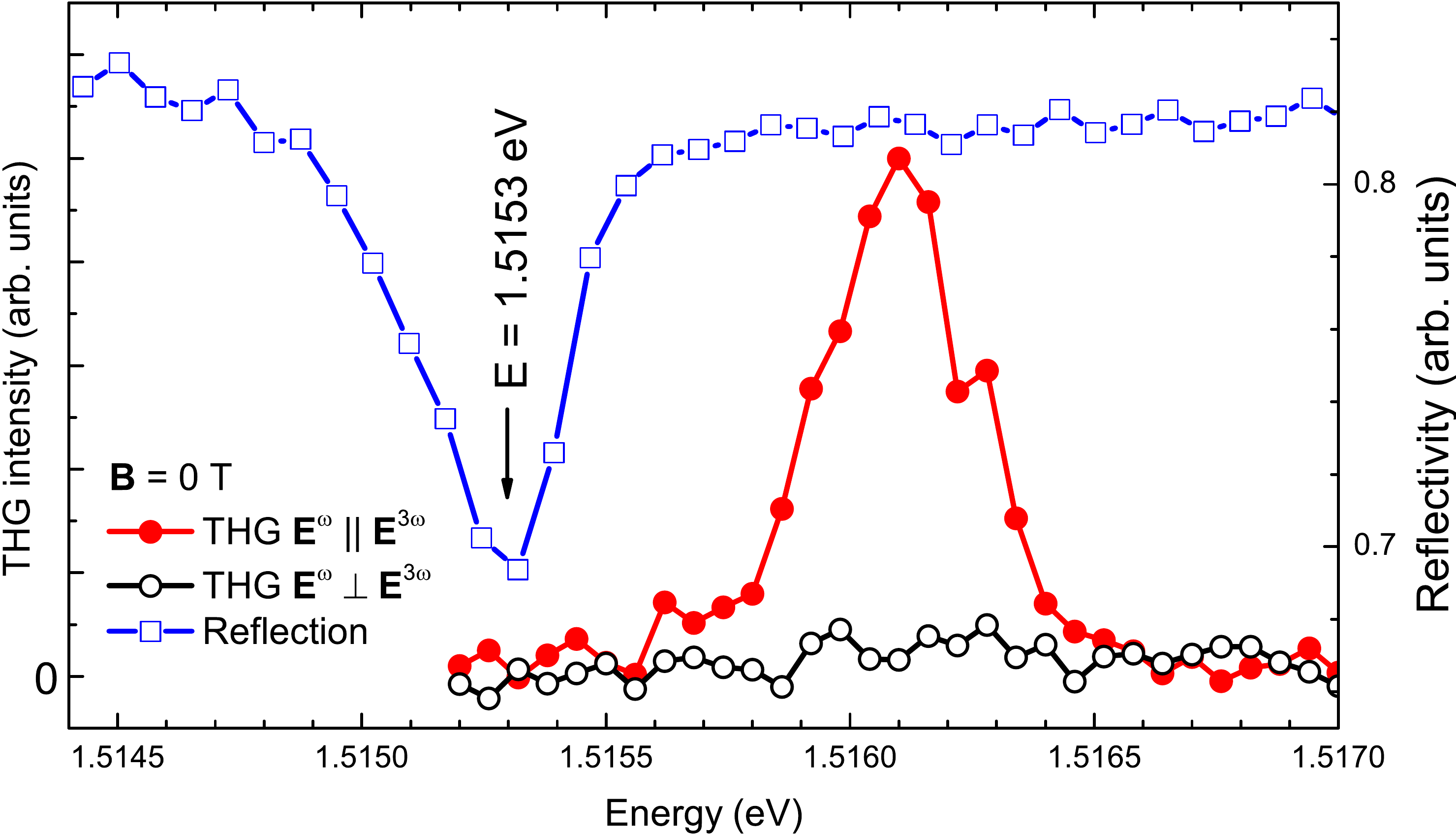}
\caption{THG and reflectivity spectra of GaAs in the vicinity of the $1s$-exciton resonance at zero magnetic field. $T=5$~K.
}
\label{fig:Fig-3}
\end{figure}

Reflectivity and THG spectra of GaAs in the vicinity of the $1s$-exciton resonance measured at zero magnetic field are shown in Fig.~\ref{fig:Fig-3}. As already noted above, the reflectivity spectrum shows a pronounced resonance of the $1s$-exciton-polariton with a minimum at 1.5153~eV, marked by the arrow, corresponding to the upper polariton branch. The THG spectrum shows a single peak with a maximum at $E_R=1.5161$~eV and a full width at half maximum of 0.3~meV. The THG peak is shifted by about $0.8$~meV to higher energy relative to the reflectivity minimum. THG signal is observed only in the parallel configuration ${\mathbf{E}^{3\omega} \parallel \mathbf{E}^{\omega}}$, while no signal is detected in the perpendicular configuration ${\mathbf{E}^{3\omega} \perp \mathbf{E}^{\omega}}$, which is in accordance to the symmetry consideration in Sec.~\ref{subsec:THG:gen:exc} below. The rotational anisotropy of the THG intensity in zero field is almost isotropic, see Fig.~\ref{fig:Fig-7}(a) and discussed further in Sec.~\ref{theory}.

\begin{figure}[t]
\centering
\includegraphics[width=0.42\textwidth]{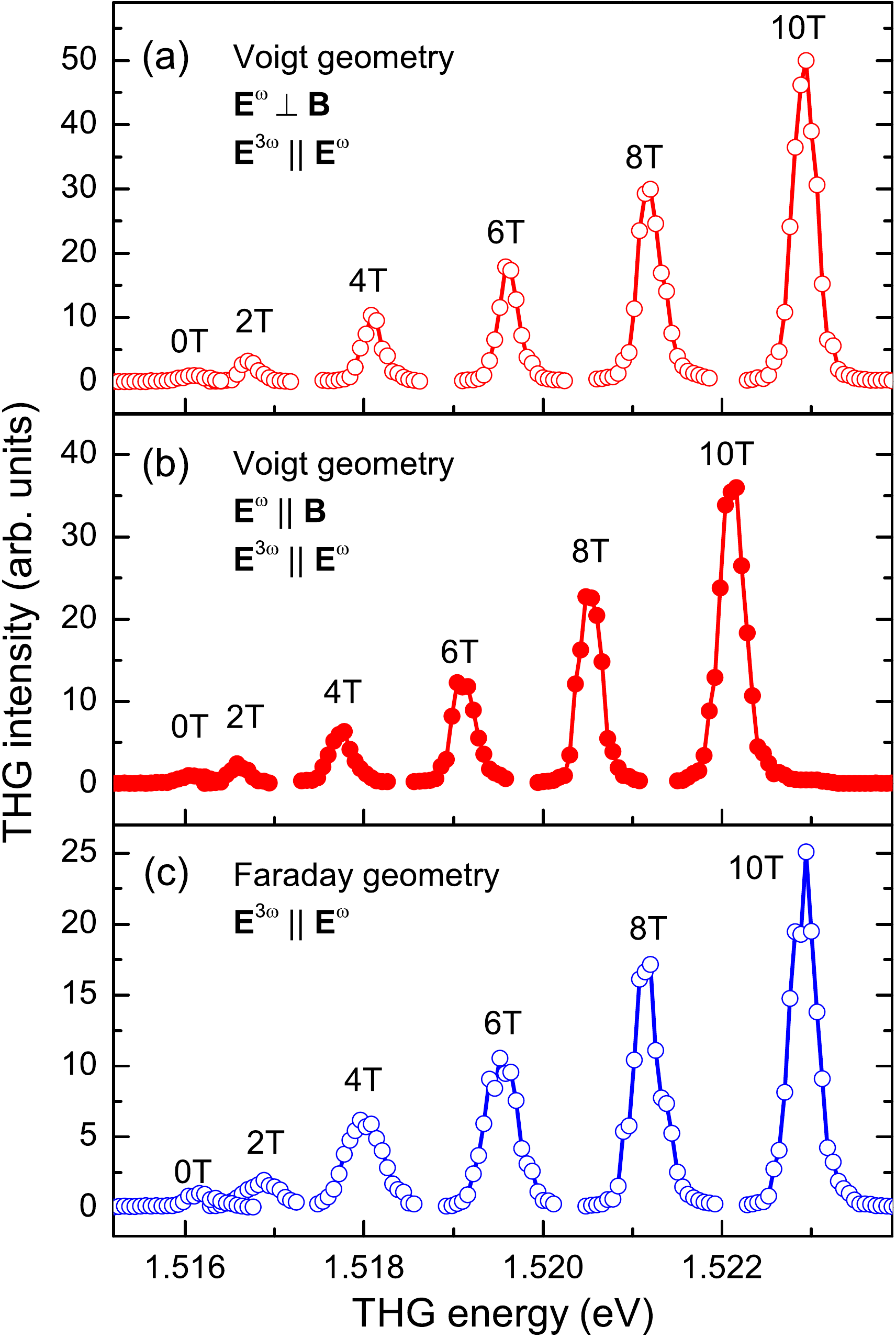}
\caption{THG spectra measured in GaAs in the vicinity of the $1s$-exciton resonance for different magnetic fields with $\mathbf{E}^{3\omega} \parallel \mathbf{E}^{\omega}$.
(a) Voigt geometry with $\mathbf{E}^{\omega} \parallel [010]$, $\mathbf{E}^{\omega} \perp \mathbf{B}$.
(b) Voigt geometry with $\mathbf{E}^{\omega} \parallel [100]$, $\mathbf{E}^{\omega} \parallel \mathbf{B}$.
(c) Faraday geometry with $\mathbf{E}^{\omega} \parallel [010]$.
}
\label{fig:Fig-4}
\end{figure}

\begin{figure}[t]
\centering
\includegraphics[width=0.45\textwidth]{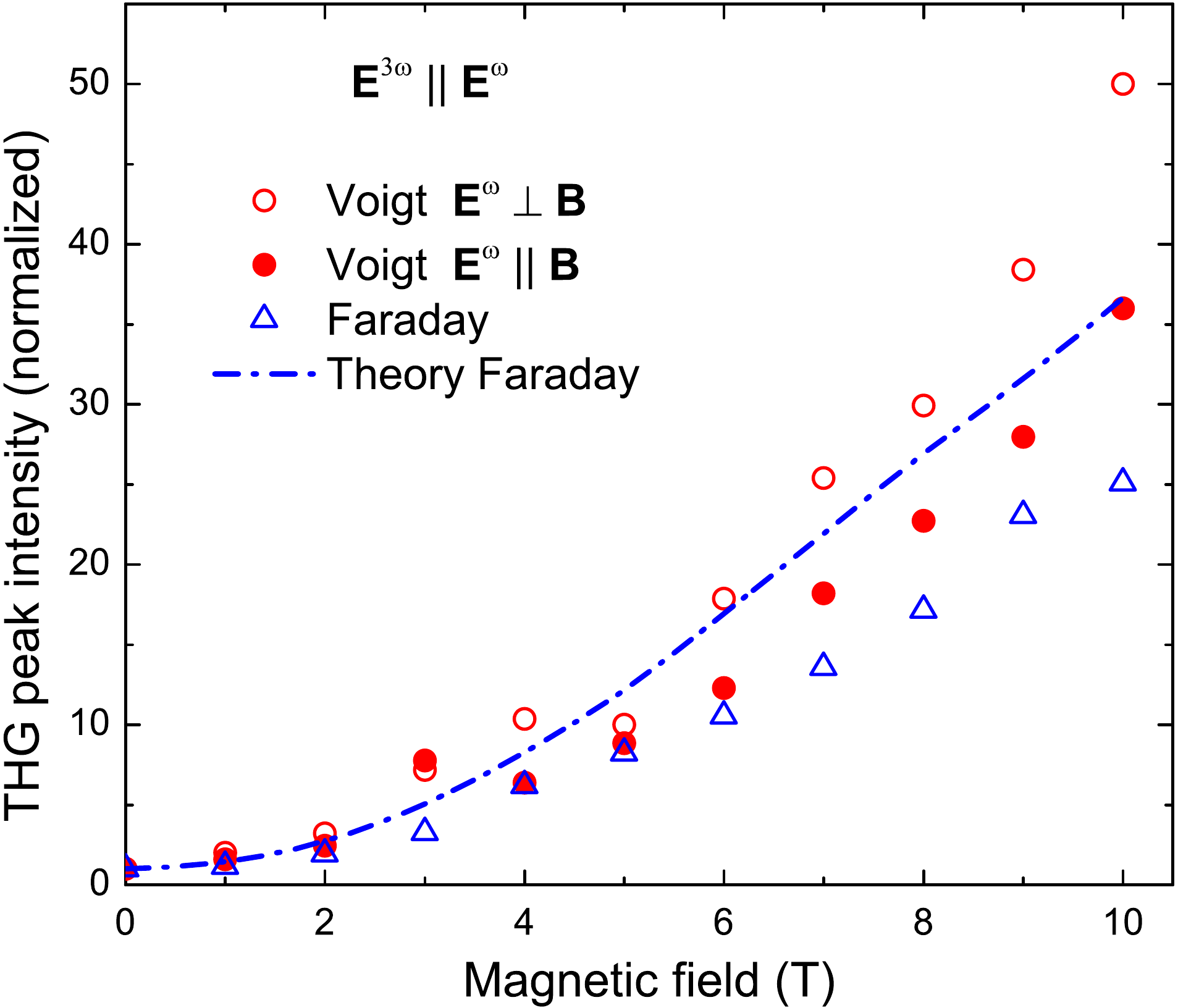}
\caption{THG intensity in GaAs as a function of magnetic field strength.
THG amplitudes are normalized on the value at $B=0$. Experimental data for
$\mathbf{E}^{3\omega} \parallel \mathbf{E}^{\omega}$ are shown by symbols.
Line is result of model calculations, see Sec.~\ref{sec:discussion} for details.
}
\label{fig:Fig-5}
\end{figure}

In external magnetic field the THG peak of the $1s$-exciton shifts to higher energies. Moreover, its intensity in the $\mathbf{E}^{3\omega} \parallel \mathbf{E}^{\omega}$ configuration drastically increases in the Voigt geometry ($\mathbf{E}^{\omega} \perp  \mathbf{B}$) with the magnetic field strength ramped from $0$ to $10$~T, see Fig.~\ref{fig:Fig-4}(a). The enhancement factor depends on the experimental geometry. The magnetic field dependencies of the THG peak intensities are shown in Fig.~\ref{fig:Fig-5}. The enhancement is 50-fold in the Voigt geometry for $\mathbf{E}^{\omega} \perp \mathbf{B}$, 36-fold in the Voigt geometry for $\mathbf{E}^{\omega} \parallel \mathbf{B}$, and 25-fold in the Faraday geometry, see Figs.~\ref{fig:Fig-4} and \ref{fig:Fig-5}. Note that the THG linewidth is nearly independent of the magnetic fields strength and, therefore, the peak intensity corresponds well to the changes of the integral THG intensity. As one can see in Fig.~\ref{fig:Fig-5}, the THG intensity grows about quadratically with magnetic field.

The strong increase of the THG intensity induced by the magnetic field is very surprising. Strong magnetic-field-induced optical harmonics generation signals were reported for the exciton resonances in GaAs, CdTe and ZnO.\cite{Saenger06,Lafrentz13,Pisarev} But these effects were observed for the SHG and in geometries where the SHG process was symmetry forbidden in zero magnetic field. The magnetic field modifies the system's symmetry and induces mechanisms for the optical harmonic generation, which result in a strong increase of the SHG intensity. The situation is fundamentally different for the THG process, as it is symmetry allowed for the studied geometries at $B=0$. In this section we present detailed experimental data for the THG signals at the exciton resonances of GaAs measured for different geometries and rotational anisotropies of the THG intensity. Reference THG data for CdTe and ZnSe are given in Sec.~\ref{THG in CdTe and ZnSe}.

One can clearly see from the comparing of the THG spectra at $B=10$~T in Figs.~\ref{fig:Fig-4}(a) and \ref{fig:Fig-4}(b) that the THG peaks in the Voigt geometry measured for the $\mathbf{E}^{\omega} \perp \mathbf{B}$ and $\mathbf{E}^{\omega} \parallel \mathbf{B}$ configurations are split in energy by about 1~meV. This splitting matches the splitting between the reflectivity peaks in these geometries shown in Fig.~\ref{fig:Fig-2}(c). The THG peak in the Faraday geometry is observed at nearly the same position as the one in the Voigt geometry, $\mathbf{E}^{\omega} \perp \mathbf{B}$, compare Figs.~\ref{fig:Fig-4}(a) and \ref{fig:Fig-4}(c).

\begin{figure}[t]
\centering
\includegraphics[width=0.45\textwidth]{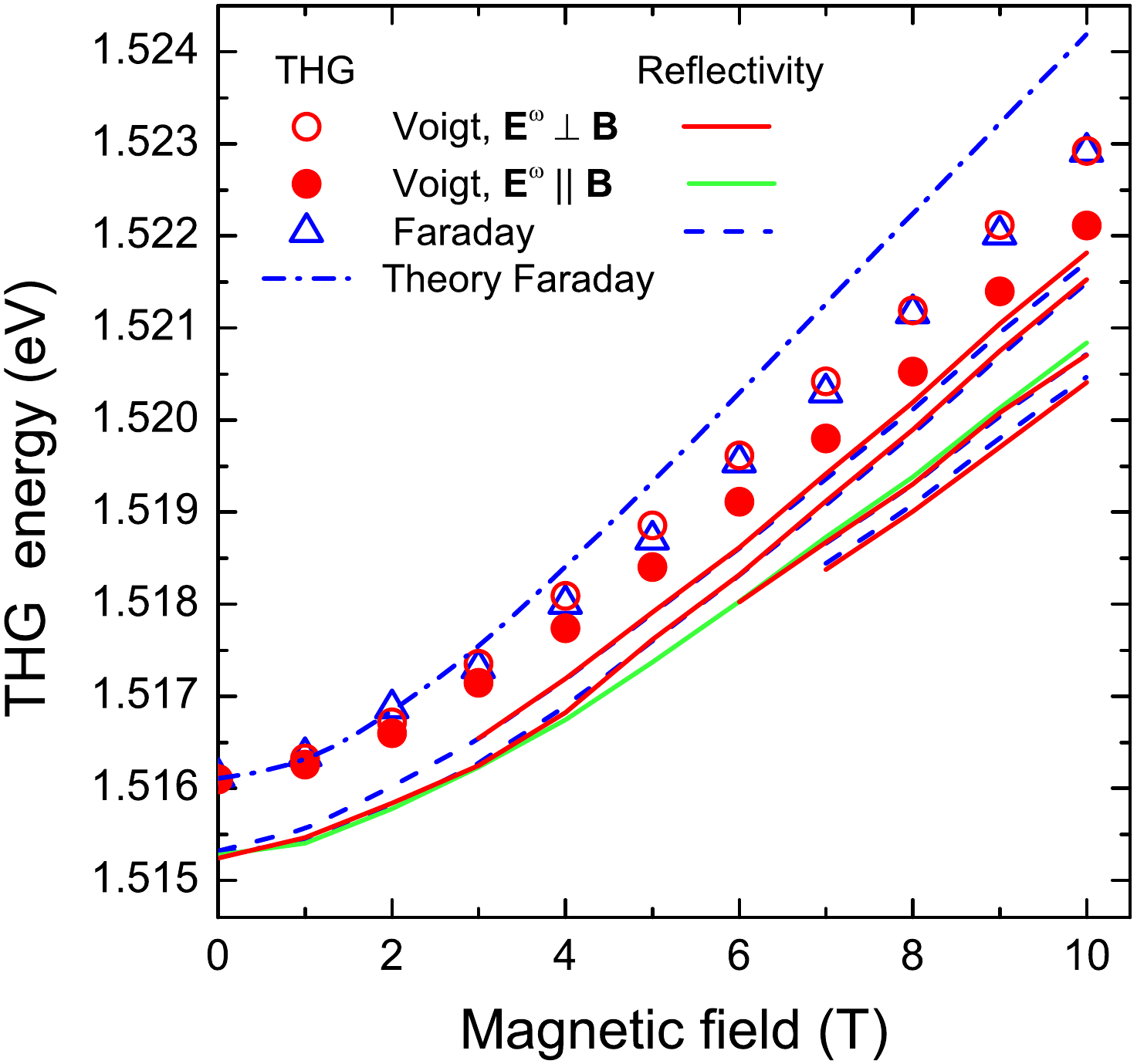}
\caption{Magnetic field dependence of exciton energies measured in THG and reflectivity for GaAs in various experimental geometries. All THG data are given for $\mathbf{E}^{3\omega} \parallel \mathbf{E}^{\omega}$. Same symbol types correspond to same data sets in Fig.~\ref{fig:Fig-5}. Symbols give THG experimental data and solid lines give reference reflectivity data from Fig.~\ref{fig:Fig-2}(c). Dashed-dotted line is result of model calculations, see Sec.~\ref{sec:discussion} for details.}
\label{fig:Fig-6}
\end{figure}

\begin{figure*}
\centering
\includegraphics[width=0.8\textwidth ,angle=0,keepaspectratio=true]{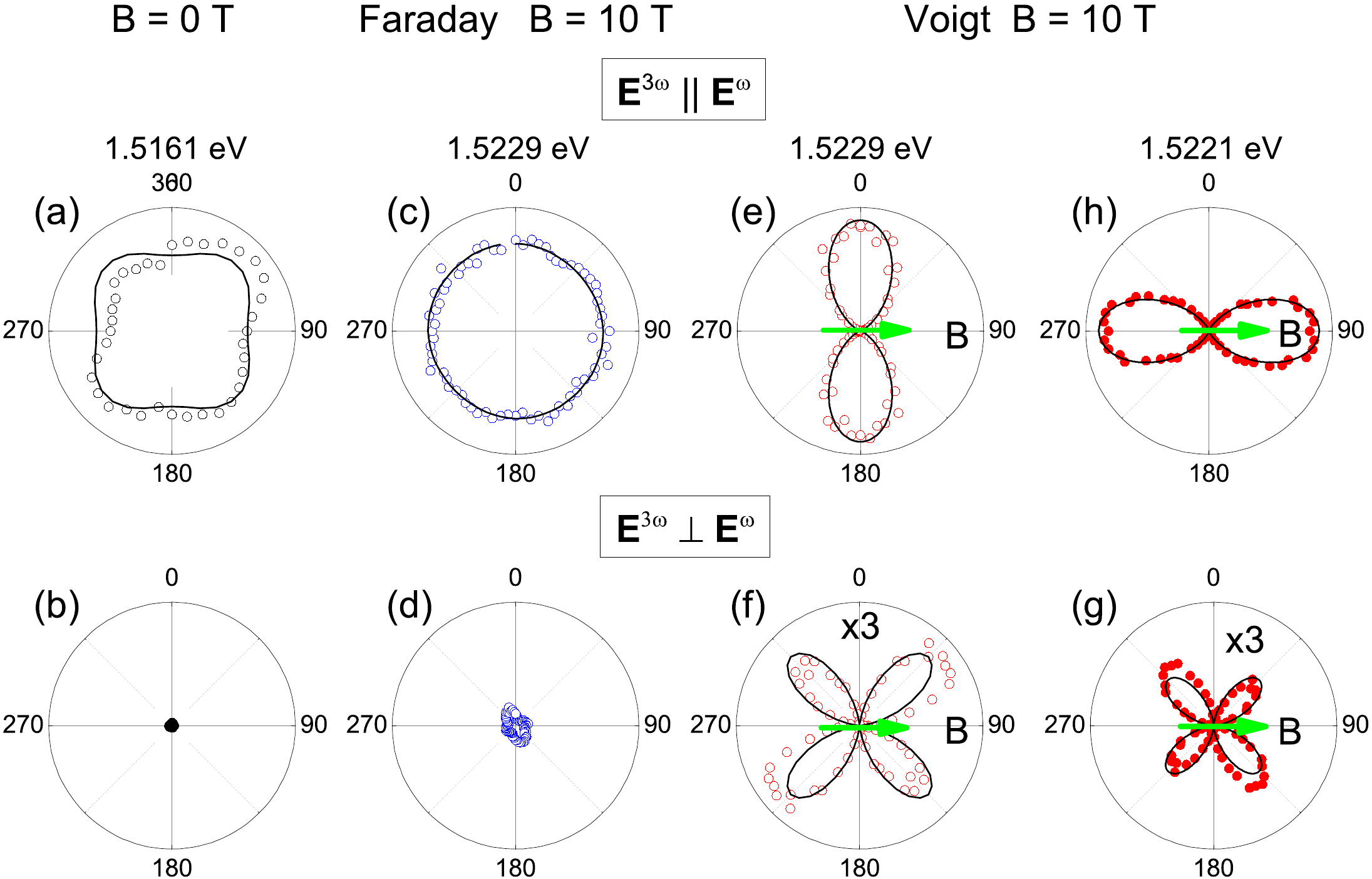}
\caption{Rotational anisotropies of the THG resonance in magnetic fields of $B=0$ and 10~T in the Voigt and Faraday geometries measured at the energies corresponding to the THG exciton peaks. Symbols are experimental data and black lines are fits discussed in Sec.~\ref{sec:discussion}. Symbol types correspond to the same data sets shown in Fig.~\ref{fig:Fig-4}.  The data in panels (a,c) are fitted with Eq.~\eqref{rotation}; in panels (e,h) with Eq.~\eqref{co} and in panels (f,g) with Eq.~\eqref{cross}.}
\label{fig:Fig-7}
\end{figure*}

The diamagnetic energy shifts of the THG lines are shown in more detail in Fig.~\ref{fig:Fig-6}. The shift in the Faraday geometry closely follows the experimental data in the Voigt geometry, $\mathbf{E}^{\omega} \perp \mathbf{B}$. For comparison with the exciton resonances in the linear optical spectra, we have added in Fig.~\ref{fig:Fig-6} reference lines mapping out the shifts of the resonances in reflectivity from  Fig.~\ref{fig:Fig-2}(c). One can see, that the reflectivity and THG data shift and split with increasing magnetic field in a very similar way. The only difference is that the THG data show the energy offset of about 1~meV, which is already present at $B=0$ and  varies only slightly with the magnetic field. The correlated behavior of the diamagnetic shifts in reflectivity and the THG peak shifts confirms that the THG signals originate from excitons. The energy shift between the reflectivity and the THG peaks points toward the polaritonic nature of the observed effect. This will be discussed in detail in Sec.~\ref{theory}.

Rotational anisotropies of the THG signal are sensitive to the crystal symmetry and magnetic-field-induced modifications of the THG selection rules and hence provide important information on the involved mechanisms of optical harmonics generation and on the involved electronic states. To record them, the THG signal is measured as a function of azimuthal polarization angle of the incoming fundamental light and the outgoing THG signal. The THG anisotropies in GaAs are shown in Fig.~\ref{fig:Fig-7} for magnetic fields of $0$ and $10$~T. In the Faraday geometry, the THG diagram has isotropic shape in the $\mathbf{E}^{3\omega} \parallel \mathbf{E}^{\omega}$ configuration [Fig.~\ref{fig:Fig-7}(c)]. Only very weak signal is detected in the $\mathbf{E}^{3\omega}\perp \mathbf{E}^{\omega}$ configuration [Fig.~\ref{fig:Fig-7}(d)]. This behavior is similar to that measured at zero magnetic field [Figs.~\ref{fig:Fig-7}(a) and \ref{fig:Fig-7}(b)].

In the Voigt geometry at $B=10$~T the THG diagrams are strongly modified. They have a two-fold rotational symmetry for the $\mathbf{E}^{3\omega} \parallel \mathbf{E}^{\omega}$ configuration, and have different orientations with respect to the magnetic field direction: perpendicular to the field for the high-energy line at 1.5229~eV [Fig.~\ref{fig:Fig-7}(e)] and parallel to it at low-energy line at 1.5221~eV [Fig.~\ref{fig:Fig-7}(h)]. It is also interesting that the THG signal appears for the $\mathbf{E}^{3\omega}\perp \mathbf{E}^{\omega}$ configuration [Figs.~\ref{fig:Fig-7}(f) and \ref{fig:Fig-7}(g)], while it is about 3 times weaker in the parallel configuration. Here, the THG diagrams demonstrate four-fold rotational symmetry with the same orientation for both exciton lines. Details on modeling the rotational anisotropies are given in Secs.~\ref{theory} and \ref{sec:discussion}.

\begin{figure}[t]
\centering
\includegraphics[width=0.48\textwidth]{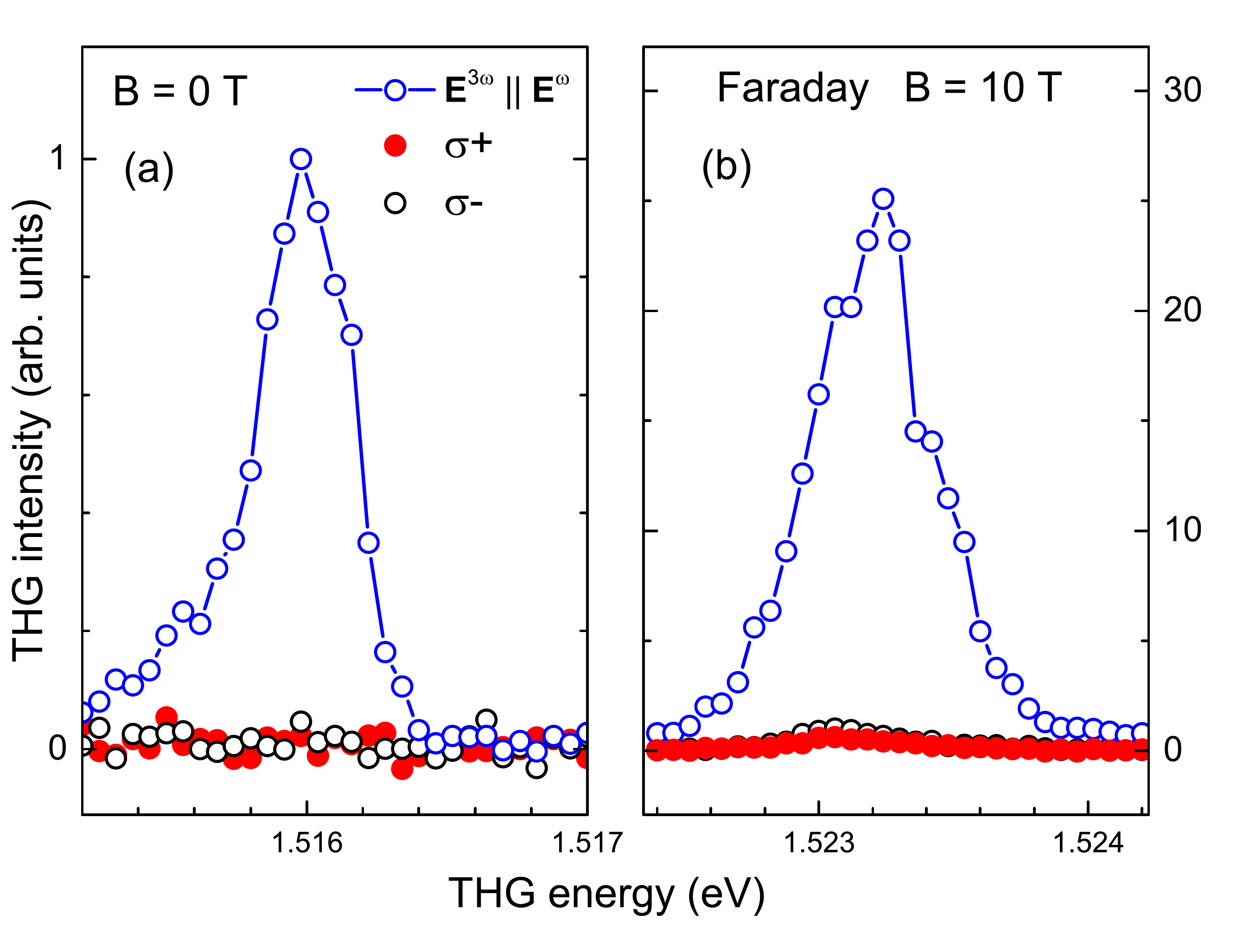}
\caption{THG spectra in GaAs at $T=5$~K, measured in magnetic fields $B=0$ and 10~T (Faraday geometry), to compare effect of linear and circular polarization of the fundamental light. THG signals are observed only for linearly polarized light (blue open circles), but absent for circularly polarized excitation (red closed and black open circles).}
\label{fig:Fig-8}
\end{figure}

We have also tested circular polarization of the excitation laser for THG generation on exciton-polaritons in GaAs. The results are shown in Fig.~\ref{fig:Fig-8}, from which one can conclude that no THG signal is visible at $B=0$. Also in magnetic field of 10~T applied in the Faraday geometry, where the THG signal induced by linearly polarized excitation is strongly enhanced, does not provide any detectable signal for circularly polarized excitation. This is due to the giant linear circular dichroism in the multiphoton absorption of cubic semiconductors\cite{Ivchenko1973} (see also Sec.~\ref{subsec:THG:gen:exc}), related with the fact that in the three photon absorption process (at $k=0$ and $B=0$) the angular momentum conservation law does not allow the absorption of three co-circularly polarized photons. Our experiment shows that, within the measurement accuracy, neither finite values of $k$ nor finite magnetic fields relax this exclusion principle.

\begin{figure}[t]
\centering
\includegraphics[width=0.45\textwidth]{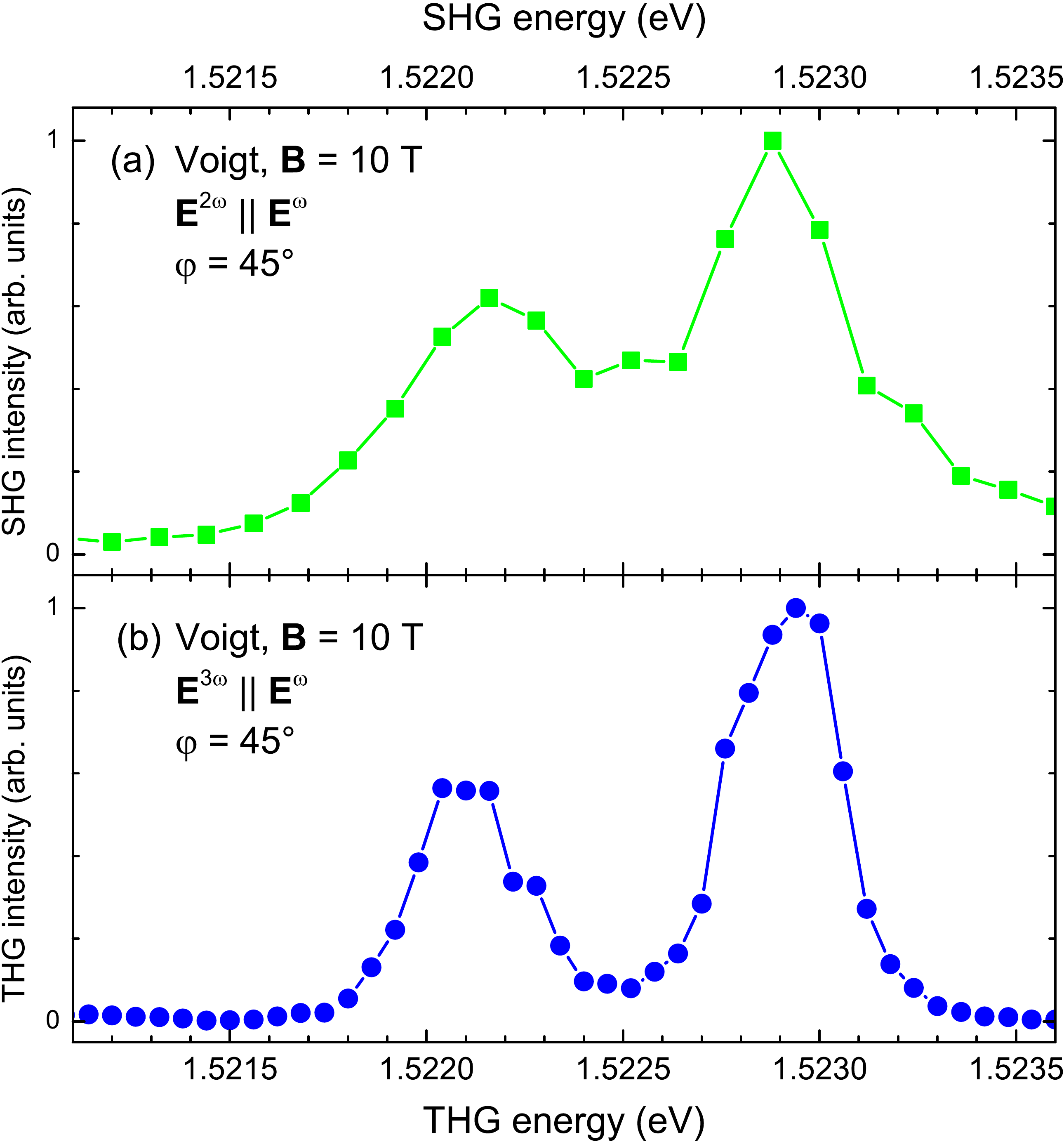}
\caption{Comparison of SHG and THG spectra in GaAs measured at $B = 10$~T in the Voigt geometry. $\varphi = 45^{\circ}$ and $T=5$~K.
}
\label{fig:Fig-9}
\end{figure}

It is instructive to compare the spectral positions of the THG and SHG signals on exciton-polaritons in GaAs. As SHG is absent for $\mathbf{k}^{\omega} \parallel [001]$ at $B=0$, we choose $B=10$~T for this comparison. The results are shown in Fig.~\ref{fig:Fig-9} for the Voigt geometry and for $\varphi = 45^{\circ}$, where both exciton polariton states are visible in the THG spectrum. One sees the same two states in the SHG spectrum with energies very close to each other. Therefore, we conclude that the exciton-polaritons can be excited by both SHG and THG. Their close energies are due to the fact that the polariton shift of the SHG and THG signals from the exciton resonance in reflectivity is rather small in GaAs, and also due to the small difference between the background refractive indices $n(E_g/2)$ and $n(E_g/3)$ at the half and the third of the band gap energy, see Sec.~\ref{theory}.

\subsection{Third harmonic generation in CdTe and ZnSe}
\label{THG in CdTe and ZnSe}

We also performed THG experiments for CdTe and ZnSe samples in magnetic field. They have the same zinc-blende crystal structure as GaAs, but have higher exciton binding energy of 10 and 20~meV in CdTe and ZnSe, respectively. Correspondingly, the exciton Bohr radius decreases from 25~nm in GaAs to 10~nm in CdTe and 6~nm in ZnSe \cite{Landolt}.
For a smaller Bohr radius the effect of the magnetic field on the exciton diamagnetic shift and the oscillator strength is reduced. We choose the same crystal orientation as shown in Fig.~\ref{fig:Fig-1}, i.e. with $\mathbf{k}^{\omega} \parallel [001]$. The magnetic field is applied in the Voigt geometry and measurements are performed in the $\mathbf{E}^{3\omega} \parallel \mathbf{E}^{\omega}$ polarization configuration.

\begin{figure}[t]
\centering
\includegraphics[width=0.5\textwidth]{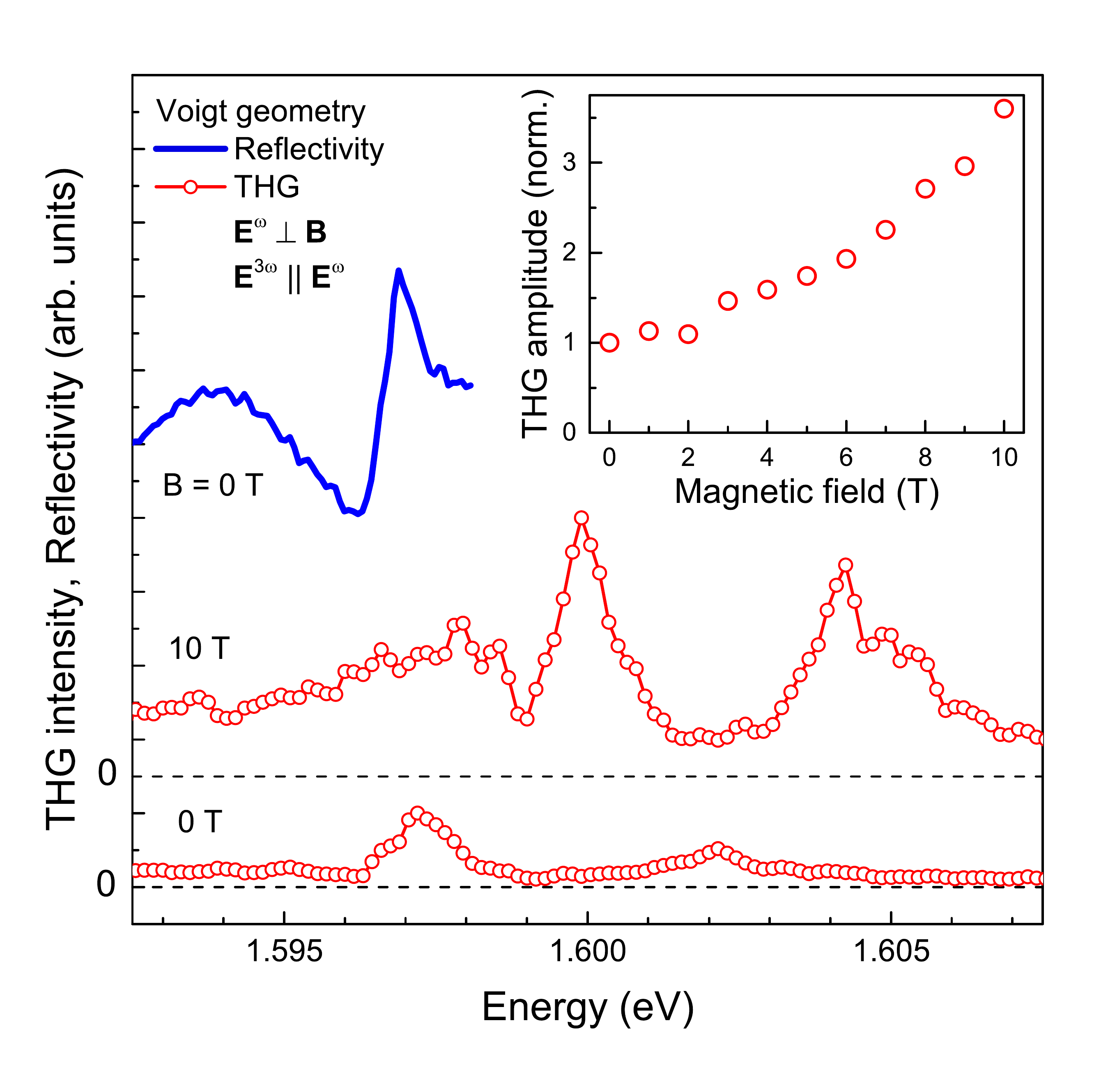}
\caption{THG and reflectivity spectra of CdTe in magnetic fields $B=0$ and 10~T at $T=5$~K. Insert shows the magnetic field dependence of the THG peak intensity for the Voigt geometry.
}
\label{fig:Fig-11}
\end{figure}

The results for CdTe are shown in Fig.~\ref{fig:Fig-11}. Here reflectivity spectrum at $B=0$, shown by the blue line, has a pronounced exciton resonance at 1.597~eV. The THG spectra are presented for $B=0$ and 10~T.  With increasing field, the exciton line shows a diamagnetic shift from 1.5972~eV to 1.600~eV and its integral intensity increases by a factor of 3.5, see the insert in Fig.~\ref{fig:Fig-11}. Similar to GaAs, the magnetic-field-induced enhancement of the THG intensity is present in CdTe, but the effect is considerably smaller, compare with Fig.~\ref{fig:Fig-5}.

\begin{figure}[t]
\centering
\includegraphics[width=0.5\textwidth]{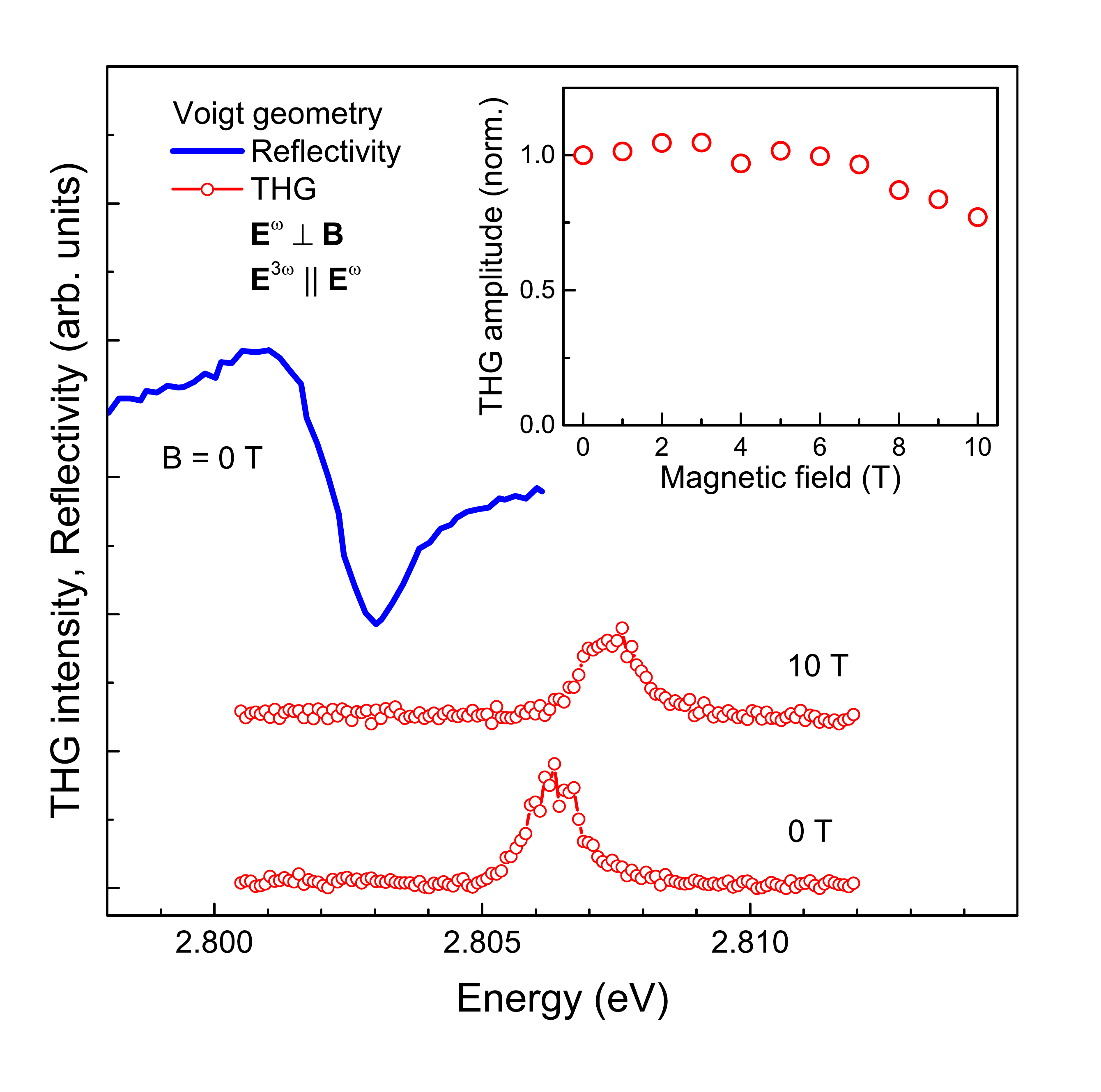}
\caption{THG and reflectivity spectra of ZnSe at zero magnetic field and 10~T. Insert shows the magnetic field dependence of THG peak intensity for the Voigt geometry.  $T=5$~K.
}
\label{fig:Fig-12}
\end{figure}

The results for ZnSe are shown in Fig.~\ref{fig:Fig-12}. The reflectivity spectrum has the exciton resonance with its minimum at 2.803~eV. The THG peak at $B=0$ is shifted to higher energy by about 3.5~meV due to the exciton-polariton effect, which is much stronger in ZnSe compared to GaAs. Due to the large exciton binding energy of 20~meV and the compact exciton size, the exciton diamagnetic shift in $B=10$~T is almost negligible. The THG integral intensity shown in inset is as well almost field independent and even demonstrates some tendency to decrease for $B$ changing from 6 to 10~T.

Comparing the results of the three studied materials we can draw a preliminary qualitative conclusion. As the strongest THG intensity increase in magnetic field is observed in GaAs, where the exciton has the largest radius and is mostly susceptible to the field, we can exclude a mechanisms not related to the exciton. We can assume that the magnetic-field-induced increase of the exciton oscillator strength is responsible for the THG intensity increase. In the next section we demonstrate that this is exactly the case by providing a theory of the THG on exciton-polaritons in semiconductors.

\section{Theory of third harmonic generation on exciton-polaritons}
\label{theory}

In this section we present the theoretical description of the third harmonic generation on exciton-polaritons in semiconductors. In Sec.~\ref{subsec:THG:gen:exc} we start with a brief outline of the general approach to calculate the third harmonic generation on excitons disregarding the spatial dispersion and magnetic field effects. In Sec.~\ref{subsec:polar} the theory of the exciton-polariton effect on the third harmonic generation is presented. Furthermore, we discuss in Sec.~\ref{subsec:fine} the manifestations of the exciton fine structure and in Sec.~\ref{subsec:magn} the magnetic field effects.

\subsection{Third harmonic generation on excitons}
\label{subsec:THG:gen:exc}

We consider the third harmonic generation in bulk cubic semiconductors at direct interband transitions. It is assumed that the incident radiation frequency $\omega$ is close to one third of the exciton resonance frequency $\omega_0 =\mathcal E_{\rm exc}/\hbar$, where $\mathcal E_{\rm exc}$ is the exciton resonance energy. The third harmonic generation can be treated as a process of three-photon excitation of exciton  followed by the coherent emission of a single photon of triple frequency $3\omega$. We take into account only transitions with virtual intermediate states in the lowest conduction and highest valence bands, i.e., the three photon absorption can be treated as a three step process $v\to c\to v\to c$. To begin with, we for simplicity  disregard the complex valence band structure of GaAs-type semiconductors. Correspondingly, the compound matrix element of the three photon absorption reads
\begin{equation}
\label{M3}
M^{(3)}E_\omega^3 = -\frac{\mathrm i e^3}{\omega^3} \frac{(\bm e\cdot\hat{\bm v})_{cv}(\bm e\cdot\hat{\bm v})_{vc}(\bm e\cdot\hat{\bm v})_{cv}}{(2\hbar\omega)^2}E_\omega^3.
\end{equation}
Here $\bm e$ is the light polarization unit vector, $\hat{\bm v}_{cv}$ is the interband  matrix element of the velocity operator, $E_\omega$ is the amplitude of the fundamental harmonic within the crystal. Hereafter the normalization volume is set to unity. Due to the significant energy deficit relative to the band gap energy at the intermediate steps of the three photon absorption~\eqref{M3} one can neglect the free carrier energy dispersion in the bands as well as the radiation wavevector.

The polarization dependence of $M^{(3)}$ can be reduced to a simple form, namely,
\begin{equation}
\label{polarization:dep}
M^{(3)}\propto (\bm e\cdot \bm e)(\bm e\cdot\hat{\bm v})_{cv}.
\end{equation}
Particularly, for circularly polarized light the three photon absorption matrix element vanishes\cite{Ivchenko1973}. This is because we consider only interband transitions and the angular momentum transferred to the crystal cannot be larger than unity for the $\Gamma_7$ to $\Gamma_6$ transition and larger than $2$ for the $\Gamma_8$ to $\Gamma_6$ transition, while three circularly polarized photons carry an angular momentum momentum of $3$. In what follows we assume that the fundamental harmonic propagates along the cubic axis $z\parallel [001]$ and the light is linearly polarized in the $(xy)$ plane. In this case one has
\begin{equation}
\label{M3:1}
M^{(3)} = -\frac{\mathrm i e^3}{\omega^3} \frac{1}{(2\hbar\omega)^2} \frac{|\tilde p_{cv}|^2\tilde p_{cv}}{m_0^3},
\end{equation}
where $\tilde p_{cv}$ is the effective interband momentum matrix element and $m_0$ is the free-electron mass. Note that for transitions from the $\Gamma_7$ to $\Gamma_6$ bands one obtains $\tilde p_{cv} = \langle \mathcal S |\hat p_x |\mathcal X\rangle/\sqrt{3}$, while neglecting the spin-orbit coupling completely (i.e., for transitions $\Gamma_{15} \to \Gamma_6$) one has $\tilde p_{cv} = p_{cv}\equiv \langle \mathcal S |\hat p_x |\mathcal X\rangle$. Here $\mathcal S$, $\mathcal X$, $\mathcal Y$, and $\mathcal Z$ are the orbital Bloch amplitudes of the electron wave functions in the conduction and valence bands, respectively.

The matrix element~\eqref{M3:1} does not depend on the electron wavevector. Hence, in agreement with the symmetry arguments, the three photon absorption is allowed at the $\Gamma$ point of the Brillouin zone and the excitation of $s$-excitons (with invariant envelope functions) is possible. We focus on the $1s$ exciton state and present the matrix element of its three photon excitation in the form
\begin{equation}
\label{exc}
M^{(3)}_{\rm exc} E_\omega^3 = \Phi_{1s}^*(0) M^{(3)} E_\omega^3,
\end{equation}
where $\Phi_{1s}(\bm r)$ is the wavefunction of the electron-hole relative motion, $\bm r$ is the relative coordinate of the electron-hole pair. In order to calculate the exciton contribution to the nonlinear dielectric polarization $\bm P^{3\omega}$ we introduce the coefficients $C_\alpha$, which describe the probability amplitudes of finding the exciton with the microscopic dipole moment $d_\alpha$ oscillating along the Cartesian axes $\alpha=x,y$, or $z$. The dielectric polarization can be recast as
\begin{equation}
\label{P:exc}
P^{3\omega}_\alpha = d C_{\alpha} + {\rm c.c.},
\end{equation}
where $d = \mathrm i e \Phi_{1s}(0) \hbar \tilde p_{cv}^*/(m_0 E_g)$ is the exciton dipole moment matrix element. Neglecting the exciton energy dispersion and the exciton-induced electromagnetic field (i.e., the polariton effect), the coefficients $C_\alpha$ can be readily expressed in the framework of time-dependent perturbation theory by means of Eq.~\eqref{exc} in the form
\begin{equation}
\label{C:alpha:no:pol}
C_\alpha = \frac{M^{(3)}_{\rm exc} E_{\alpha,\omega}^3}{\mathcal E_{\rm exc} - 3\hbar\omega}.
\end{equation}

It is convenient to describe the nonlinear response of the crystal in the absence of polariton effects by the forth-rank tensor  $\chi_{\alpha\beta\gamma\delta}$ with $\alpha, \ldots, \delta$ being the Cartesian axes which describes the third-order nonlinear susceptibility as
\begin{equation}
\label{phenom:simple}
P^{3\omega}_\alpha = \chi_{\alpha\beta\gamma\delta} E_{\beta,\omega} E_{\gamma,\omega} E_{\delta,\omega}.
\end{equation}
The tensor $\chi_{\alpha\beta\gamma\delta}$ is symmetric to any permutation of the last three indices. In the $T_d$ point symmetry group relevant for the studied samples there are two linearly independent components
\begin{subequations}
\label{phenom:simple:1}
\begin{align}
&\chi_{\alpha\alpha\beta\beta} = \chi_{\alpha\beta\beta\alpha} = \chi_{\alpha\beta\alpha\beta} = \mathcal A, \quad \alpha\ne\beta, \\
&\chi_{\alpha\alpha\alpha\alpha} = \mathcal C.
\end{align}
\end{subequations}

Based on the analysis of Eq. \eqref{phenom:simple}, one can write the rotation anisotropies for the THG intensities  in the parallel, $I^{3\omega}_{\parallel}$ for ${\bm E}^{3\omega} \parallel {\bm E}^{\omega}$, and perpendicular,  $I^{3\omega}_{\perp}$ for ${\bm E}^{3\omega} \perp {\bm E}^{\omega}$, polarization geometries  as
	\begin{eqnarray} \label{rotation}
	&&	I^{3\omega}_{\parallel}(\varphi) \propto  \left|\mathcal C + \frac{1}{2}(\mathcal A-\mathcal C)\sin^2{2\varphi}\right|^2 \, , \\
	&& I^{3\omega}_{\perp}(\varphi) \propto  \frac{1}{16}|(\mathcal A - \mathcal C)\sin{4\varphi}|^2  \nonumber\, ,
	\end{eqnarray}
	where $\varphi$  is defined in Fig.~\ref{fig:Fig-1}.
It follows from Eqs.~\eqref{polarization:dep} and \eqref{C:alpha:no:pol} that in the considered model where the intraband transitions (as well as the contributions of remote bands) are disregarded, the constants $\mathcal A$ and $\mathcal C$ are equal. Making use of Eq.~\eqref{exc} we obtain analytically
\begin{equation}
\label{chi3:exc}
\mathcal A = \mathcal C = \frac{e^4}{\omega^3}\frac{\hbar |\tilde p_{cv}|^4|\Phi_{1s}(0)|^2}{m_0^4 E_g (\hbar\omega)^2(\mathcal E_{\rm exc} - 3\hbar\omega)} {\propto \frac{|\Phi_{1s}(0)|^2}{\mathcal E_{\rm exc} - 3\hbar\omega}}.
\end{equation}
The susceptibility, as expected, has a resonance at $3\hbar\omega = \mathcal E_{\rm exc}$ and is proportional to the propability to find the electron and hole within the same unit cell, $|\Phi_{1s}(0)|^2$. One can see, that the prediction $\mathcal A=\mathcal C$ in our model results into THG, $I^{3\omega}_{\parallel}=|\mathcal C|^2$, for the parallel geometry and vanishing THG, $I^{3\omega}_{\perp}=0$, for the perpendicular geometry. The polariton effects which we consider in the following subsection do not affect these conclusions.

\subsection{Polariton effects}\label{subsec:polar}

\begin{figure}[t]
\centering
\includegraphics[width=0.4\textwidth,height=0.35\textwidth]{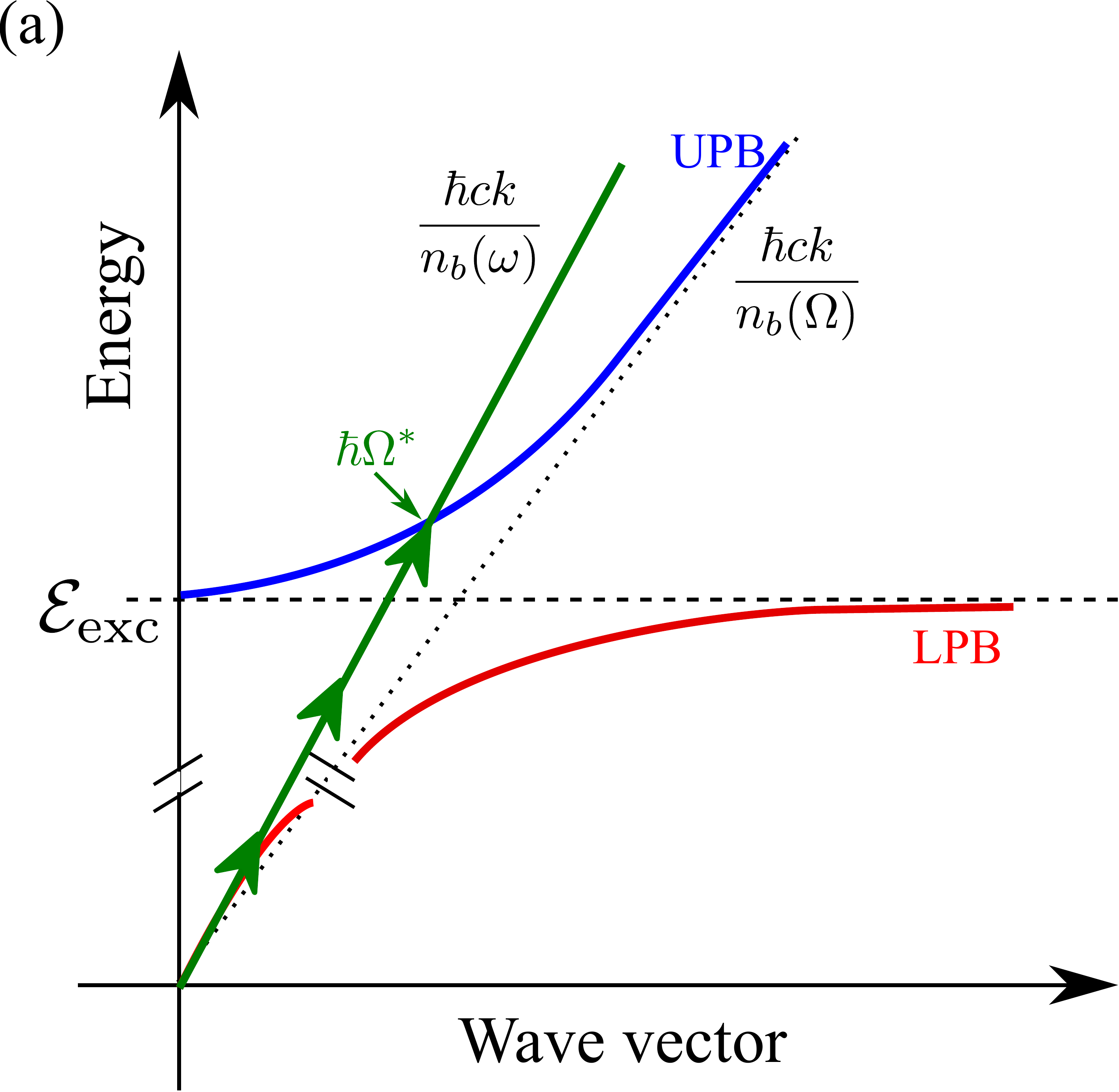}
\includegraphics[width=0.4\textwidth,height=0.35\textwidth]{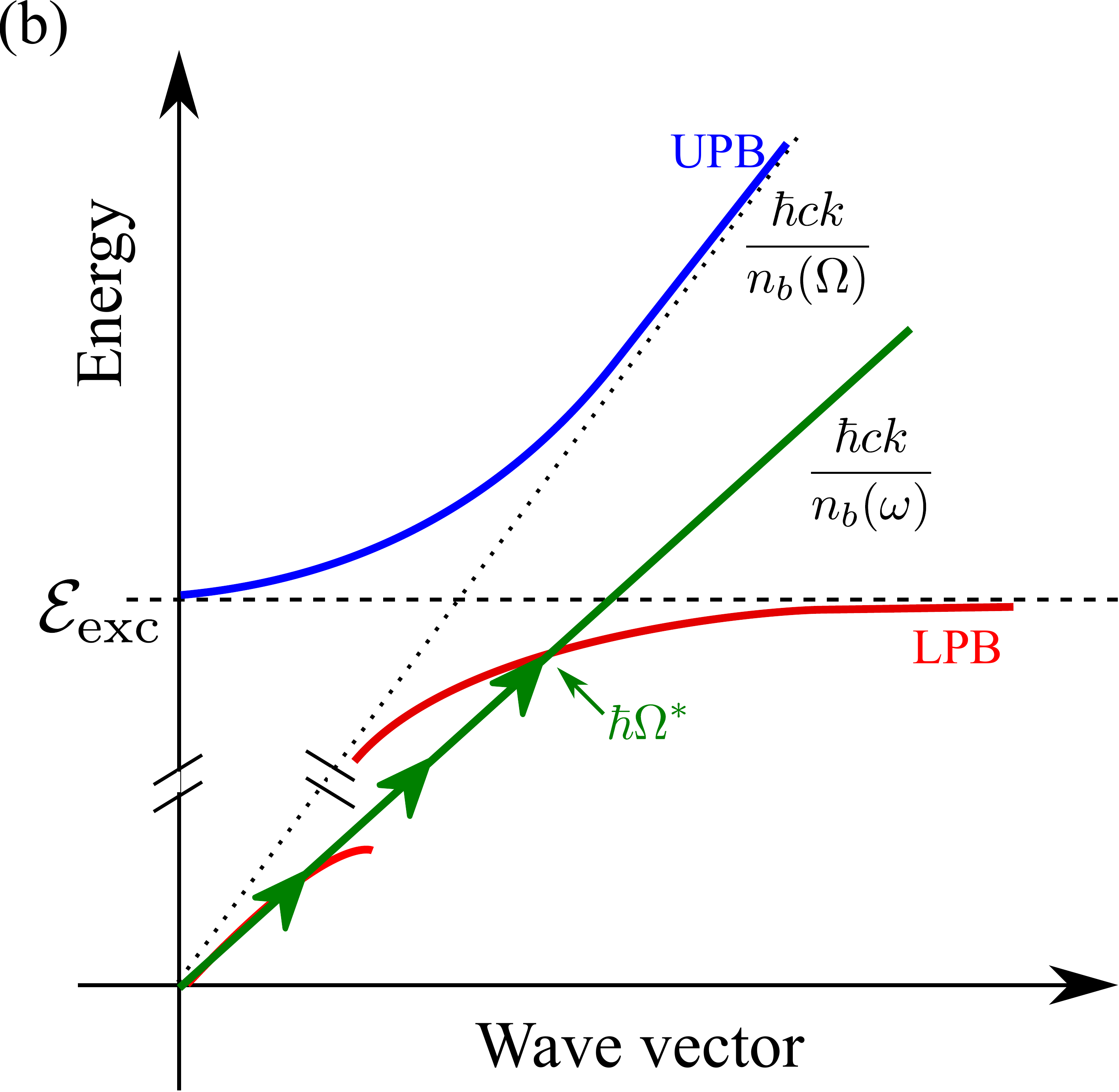}
	\caption{(a) and (b) Schematics of the three-photon excitation process of exciton-polaritons contributing to the THG for $n_b(\omega)<n_b(\Omega)$  and $n_b(\omega)>n_b(\Omega)$, respectively.
		}
\label{fig:Fig-13}
\end{figure}

In bulk semiconductors the light-matter interaction strongly modifies the energy spectrum of the excitons and affects the propagation of the radiation via the formation of exciton-polaritons. This is particularly important for the excitonic mechanism of the THG because the frequency of the third harmonic, $3\omega$, is in the vicinity of excitonic resonances. This requires an extension of the approach of Sec.~\ref{subsec:THG:gen:exc} to account for the polariton effect. In this subsection, for simplicity, we consider the transitions between the bands $\Gamma_7 \to \Gamma_6$  where the exciton is characterized by a single translational mass $M$. In this case the polariton dispersion curve consists of two branches, the lower (LPB) and upper (UPB) one, see Figs.~\ref{fig:Fig-13}(a) and \ref{fig:Fig-13}(b). The more complex case of $\Gamma_8\to \Gamma_6$ is considered in Sec.~\ref{subsec:fine}, see also Fig.~\ref{fig:Fig-14}.

To that end we present the nonlinear polarization $\bm P^{3\omega}$ and the induced electric field $\bm E^{3\omega} \equiv \bm E_{\Omega}$  at the third harmonic frequency $\Omega = 3\omega$ as a sum of a particular solution of the inhomogeneous set of Schr\"{o}dinger and Maxwell equations and the general solution of the homogeneous equations in order to account for the boundary conditions at the crystal surfaces. First, we address the inhomogeneous solution which describes the THG in the infinite crystal. The coefficients $C_\alpha$ and the electric field $E_{\alpha,\Omega}^+$ oscillating as $\exp({-\mathrm i\Omega t})$ (the so-called positive frequency term) satisfy the coupled system of equations
\begin{subequations}
\label{polariton}
\begin{equation}
\label{C:eq}
 \left( \mathcal E_{\rm exc} + \frac{\hbar^2 K^2}{2M} - \hbar \Omega - \mathrm i \hbar\gamma \right) C_\alpha  = dE_{\alpha,\Omega}^{(+)}+ M^{(3)}_{\rm exc} E_{\alpha,\omega}^3,
\end{equation}
\begin{equation}
\label{E:eq:+}
\left[\frac{\Omega}{c}n_b(\Omega) - K\right] E^+_{\alpha,\Omega} = -\frac{2\pi}{n_b(\Omega)} \frac{\Omega}{c} dC_{\alpha}.
\end{equation}
\end{subequations}
Here $c$ is the speed of light, $M$ is the translational exciton mass, $\bm K$ is the exciton wavevector equal to $3\bm k^\omega$, $\bm k^\omega$ is the wavevector of the incident radiation, i.e., of the fundamental harmonic, $n_b\equiv n_b(\Omega)$ is the  background refraction index that is weakly dependent on the frequency and related with the background dielectric constant $\varepsilon_b{(\Omega)}$ found neglecting the $1s$ exciton resonance, $\hbar\gamma$ is the nonradiative damping rate of the exciton. In what follows we assume that the difference of the refractive indices at the fundamental and third harmonic frequencies is small, $|n_b(\omega)-n_b(\Omega)| \ll n_b(\omega)$.

We introduce the bare-exciton Green's function
\begin{equation}
\label{G:exc}
G(\Omega, K) = \frac{1}{\hbar \Omega  -\mathcal E_{\rm exc} - \frac{\hbar^2 K^2}{2M}+\mathrm i \hbar\gamma},
\end{equation}
which allows us to present the solution of the set~\eqref{polariton} in the compact form
\begin{equation}
\label{E:eq:+:1}
E_{\alpha,\Omega}^+ = \frac{2\pi}{n_b(\Omega)} dM^{(3)}_{\rm exc} E_{\alpha,\omega}^3 \frac{G(\Omega,K)}{n_b(\Omega) - \frac{cK}{\Omega} - \frac{2\pi}{n_b(\Omega)} d^2 G(\Omega,K)}.
\end{equation}
It is noteworthy that the poles of Eq.~\eqref{E:eq:+:1} correspond to the transversal polariton modes:
\begin{equation}
\label{disper:resonant}
\varepsilon(\Omega, K) = \left(\frac{cK}{\Omega} \right)^2=n_b^2(\omega),
\end{equation}
where the dielectric susceptibility with account for the exciton resonance reads in the resonant approximation
\begin{equation}
\label{epsilon:exc}
\varepsilon(\Omega, K) = \varepsilon_b(\Omega) + \frac{4\pi d^2}{\mathcal E_{\rm exc} + \frac{\hbar^2 K^2}{2M} - \hbar\Omega - \mathrm i \hbar\gamma}.
\end{equation}
The numerator in Eq.~\eqref{epsilon:exc} can be conveniently recast as
\begin{equation}
\label{osc:str}
4\pi d^2 = \hbar\omega_{LT} \varepsilon_b(\mathcal E_{\rm exc}/\hbar) \propto |\Phi_{1s}(0)|^2,
\end{equation}
where $\omega_{LT}$ is the exciton longitudinal-transverse splitting.

In order to analyze the spectral dependence of the RHG effect we substitute the function~\eqref{G:exc} into Eq.~\eqref{E:eq:+:1} and obtain in the vicinity of $3\hbar\omega \approx \mathcal E_{\rm exc}$
\begin{equation}
\label{THG:resonance}
E^+_{\alpha,\Omega} \propto \frac{{|\Phi_{1s}(0)|^2}}{n_b(\Omega) - n_b(\omega)} \frac{E_{\alpha,\omega}^3}{3\hbar\omega - \hbar\Omega^*(3q)},
\end{equation}
where
\begin{multline}
\label{Omega*}
\hbar\Omega^*(K) =\\
 \mathcal E_{\rm exc} + \frac{\hbar^2 K^2}{2M} + \frac{\hbar\omega_{LT}}{2} \frac{n_b(\mathcal E_{\rm exc}/\hbar)}{n_b(\Omega) - n_b(\omega) - \mathrm i \hbar \gamma}.
\end{multline}
Equations~\eqref{THG:resonance} and \eqref{Omega*} demonstrate that the field of the third harmonic as a function of the fundamental frequency has a resonance at
\begin{equation}
3\omega = \Re\{\Omega^*(3k^\omega)\}.
\end{equation}
It follows from Eq.~\eqref{Omega*} that the THG resonance is shifted from the exciton resonance frequency $\mathcal E_{\rm exc}/\hbar$ due to (i) the mechanical exciton dispersion, $\hbar^2K^2/2M$ and (ii) the polariton effect described by the second term in Eq.~\eqref{Omega*}. The interpretation of the polariton shift immediately follows from the conservation laws of energy and momentum, which are fulfilled in the third order nonlinear process.  In fact, the THG is most efficient when the energy of the three photons $3\omega$ matches the polariton energy at the wavevector $3\bm k^\omega$, see Fig.~\ref{fig:Fig-13}. Indeed, making use of the dispersion equation~\eqref{disper:resonant} and noting that at the resonance $cK/\Omega = ck^\omega/\omega = n_b(\omega)$ we obtain Eq.~\eqref{Omega*}. For a large value of $M$ the physical root of this equation reads
\[
\omega = \Omega^*(3k^\omega)/3.
\]
Another root corresponds to an unrealistically high energy and is unphysical. Noteworthy, in the three-photon resonant absorption, an exciton-polariton is generated at the upper polariton branch, UPB, if $n_b(\omega)<n_b(\Omega)$ [Fig.~\ref{fig:Fig-13}(a)], and at the lower branch, LPB, if $n_b(\omega)>n_b(\Omega)$ [Fig.~\ref{fig:Fig-13}(b)]. We stress that the polariton effect allows one to fulfill the phase matching condition and results in the strong enhancement of the THG signal at the exciton-polariton resonances.

Interestingly, if $n_b(\omega)=n_b(\Omega)$, e.g., if the frequency dispersion of the background dielectric constant is absent, then the exciton resonance in the third harmonic vanishes, namely,
\begin{equation}
\label{no:disper}
E_{\alpha,\Omega}^+ =-\frac{M^{(3)}_{\rm exc} E_{\alpha,\omega}^3}{d}.
\end{equation}
In this regime of full synchronization the left hand side of Eq.~\eqref{E:eq:+} vanishes, the coefficient $C_\alpha$ and, hence, the left hand side of Eq.~\eqref{C:eq} vanish as well and we immediately obtain Eq.~\eqref{no:disper}. In fact, this means that the light dispersion crosses neither the upper nor the lower polariton branches.

Now we briefly analyze the propagation of the third harmonic light in the bounded crystal slab. To illustrate the most important effects it is sufficient to consider a semi-infinite crystal occupying the half-space $z>0$. Let the fundamental harmonic be incident at the $z=0$ boundary. Since the frequency $\omega$ lies in the transparency region, there are no additional light waves at this frequency and, moreover, we can disregard the absorption of the fundamental harmonic. Thus, the incident field inside the crystal is given by
\[
E_{\alpha,\omega} e^{-\mathrm i \omega t+ \mathrm i k^\omega z} + {\rm c.c.},
\]
where the amplitude $E_{\alpha,\omega}$ of the transmitted light is found from the standard Fresnel boundary conditions.
The field of the third harmonic consists of three contributions, the solution Eq.~\eqref{E:eq:+:1} or \eqref{THG:resonance} of the inhomogeneous set~\eqref{polariton}, and two polariton waves satisfying the homogeneous set~\eqref{polariton}, as follows
\begin{equation}
\label{field:to:right}
E_{\alpha,\Omega}(z,t)=\left(E_{\alpha,\Omega}^+ e^{\mathrm i K z} + E_{\alpha,\Omega}^{(1)}e^{\mathrm i K_1 z}+ E_{\alpha,\Omega}^{(2)}e^{\mathrm i K_2 z}\right)e^{-\mathrm i \Omega t}.
\end{equation}
Here the wavevectors $K_{1,2}$ are determined from Eq.~\eqref{disper:resonant} and $K_1$ is chosen to be closest to the $3k^\omega$. The fields $E_{\alpha,\Omega}^{(1)}$ and $E_{\alpha,\Omega}^{(2)}$ can be found from the boundary conditions for the electric field, which require continuity of the tangential components of the electric field and the normal components of the magnetic field, namely,
\begin{subequations}
\label{BC0}
\begin{align}
E_r &= E_{\alpha,\Omega}^+ +  E_{\alpha,\Omega}^{(1)} + E_{\alpha,\Omega}^{(2)},\\
-\frac{\Omega}{c}E_r &= KE_{\alpha,\Omega}^+ + K_1  E_{\alpha,\Omega}^{(1)} + K_2 E_{\alpha,\Omega}^{(2)},
\end{align}
\end{subequations}
where $E_r$ is the third harmonic generated at the boundary and propagating out of the crystal towards $z\to - \infty$. The third boundary condition required because of the spatial dispersion
\begin{multline}
\label{abc}
0= \frac{d E_{\alpha,\Omega}^{(1)}}{\hbar\Omega - \mathcal E_{\rm exc} - \frac{\hbar^2 K_1^2}{2M}}
+ \frac{d E_{\alpha,\Omega}^{(2)} }{\hbar\Omega - \mathcal E_{\rm exc} - \frac{\hbar^2 K_2^2}{2M}} \\
+ \frac{d E_{\alpha,\Omega}^+ - M^{(3)}_{\rm exc} E_{\alpha,\omega}^3}{\hbar\Omega - \mathcal E_{\rm exc} - \frac{\hbar^2 K^2}{2M}}
\end{multline}
is the Pekar condition for the exciton polarization vanishing at the boundary.

Equations~\eqref{BC0} and \eqref{abc} allow us to calculate the fields within the crystal. For the typical parameters the reflected field $E_r$ is small and, instead of the two conditions~\eqref{BC0}, one can use the requirement
\begin{equation}
0 = E_{\alpha,\Omega}^+ +  E_{\alpha,\Omega}^{(1)} + E_{\alpha,\Omega}^{(2)},\label{BC:E0}
\end{equation}
which states that the field of the total third harmonic is zero at $z=0$. Making use of Eqs.~\eqref{BC:E0} and \eqref{abc} we obtain
\begin{widetext}
\begin{multline}
\label{to:right}
E_{\alpha,\Omega}(z,t) = E_{\alpha,\Omega}^+e^{\mathrm i Kz- \mathrm i \Omega t} - \frac{ E_{\alpha,\Omega}^+ [G(\Omega,K) - G(\Omega,K_2)] + (M^{(3)}_{\rm exc}E_{\alpha,\omega}^3/d)G(\Omega,K)}{G(\Omega,K_1) - G(\Omega,K_2)} e^{\mathrm i K_1z- \mathrm i \Omega t} -\\
 \frac{ E_{\alpha,\Omega}^+ [G(\Omega,K_1) - G(\Omega,K)] - (M^{(3)}_{\rm exc}E_{\alpha,\omega}^3/d)G(\Omega,K)}{G(\Omega,K_1) - G(\Omega,K_2)} e^{\mathrm i K_2z- \mathrm i \Omega t}.
\end{multline}
\end{widetext}
Equation~\eqref{to:right} generalizes the standard expressions for the harmonic generation in crystals~\cite{Bloembergen} to account for the exciton-polariton effects.

For the heavy translational mass $M$, where the spatial dispersion is unimportant, but the frequency dispersion of the background dielectric constant can be important, the wave with the wavevector $K_2$ is not excited, Eq.~\eqref{to:right} can be reduced to
\begin{multline}
\label{no:space:1}
E_{\alpha,\Omega}(z,t) = \frac{2\pi}{n_b(\Omega)} dM_{exc}^{(3)} E_{\alpha,\omega}^3 e^{-\mathrm i \Omega t}\\
\times \frac{G(\Omega,0)}{n_b(\Omega) - n_b(\omega) - \frac{2\pi}{n_b(\Omega)} d^2 G(\Omega, 0)}\left(e^{\mathrm i Kz}-e^{\mathrm i K_1z}\right) \\
{\propto \frac{|\Phi_{1s}(0)|^2 \left(e^{\mathrm i Kz}-e^{\mathrm i K_1z}\right)}{K - K_1}}.
\end{multline}
Generally, the wavevectors {of the polariton} $K_1 = 3k^\omega \sqrt{{\varepsilon(3\omega)}}/n_b(\omega)$  and $K=3k^\omega$ are not equal, thus the intensity of the third harmonic inside the crystal, $I^{3\omega} \propto \sin^2{[(K-K_1)z]}$  oscillates as a function of coordinate. This is because the phase matching condition is not fulfilled at {$3\omega \ne \Omega^*$ in Eq.~\eqref{Omega*}, i.e., where} $K_1\ne K$. Exactly at the THG resonance (neglecting damping) $K_1=K$ and the denominator in Eq.~\eqref{no:space:1} vanishes, which corresponds to the phase-matching condition due to the exciton-polariton. Hence, one has to consider the limit of $3\omega \to \Omega^*$ and making use of the l'Hospital rule one obtains $E_{\alpha}(z,t) \propto z$ and the THG intensity $I^{3\omega}\propto |E_{\alpha,\Omega}(z,t)|^2$ increasing quadratically with the coordinate.

\subsection{Role of exciton fine structure}
\label{subsec:fine}

Let us now analyze the modifications of Eqs.~\eqref{M3:1} and \eqref{exc} with allowance for the complex valence band structure. In zinc-blende structure the complex valence band is formed of the  $\Gamma_8$ and the spin-orbit split $\Gamma_7$ subbands, while the conduction band bottom transforms according to the $\Gamma_6$ spinor representation. The $s$-exciton states transform according to the reducible representation
\begin{equation}
\label{decomposition}
\Gamma_6 \otimes \Gamma_8 =  E \oplus F_1 \oplus F_2,
\end{equation}
where the states belonging to the $E$ and $F_1$ irreducible representation are dark, while the three states belonging to the $F_2$ representation are optically active in the dipole approximation.
The intermediate states for the $\Gamma_8\to \Gamma_6$ three-photon transition can be both in the same $\Gamma_8$ or in the split-off $\Gamma_7$ band. The matrix element of the three photon excitation of the optically active states reads
\begin{multline}
\label{M3:complex}
M_{\rm exc}^{(3)}E_{\alpha,\omega}^3 = -\frac{\mathrm i \Phi^*_{1s}(0) e^3}{\omega^3} \frac{|p_{cv}|^2 p_{cv}}{3m_0^3}  \\
\times\left[\frac{2}{(2\hbar\omega)^2} + \frac{1}{2\hbar\omega(2\hbar\omega+\Delta_{so})}\right]E_{\alpha,\omega}^3,
\end{multline}
where $\Delta_{so}$ is the valence band spin-orbit splitting, $p_{cv} = \langle \mathcal S |\hat{p}_x |\mathcal X\rangle$. Accounting for the valence band structure does not change the polarization dependence of the matrix element and the relation between the components of the susceptibility tensor $\mathcal A=\mathcal C$ in Eqs.~\eqref{phenom:simple:1}.  The mixing of the exciton states $E$ and $F_1$ with optically active states $F_2$ due to the non-zero exciton wavevector and the external magnetic field is discussed in Sec.~\ref{subsec:magn}.

In Secs.~\ref{subsec:THG:gen:exc} and \ref{subsec:polar} the model of the THG involved only three optically active exciton states transforming according to the irreducible representation $F_2$ of the $T_d$ point symmetry group, see Eq.~\eqref{decomposition}. The fine structure of the $\Gamma_6\otimes \Gamma_8$ exciton at rest ($K=0$) is controlled by the short-range exchange interaction: Its isotropic part splits otherwise 8-fold degenerate exciton state into the triplet $F_2$ corresponding to the angular momentum $F=1$ and the 5-fold degenerate state with $F=2$. Due to the cubic anisotropy the $F=2$ state splits into a doublet and a triplet transforming according to the representations $E$ and $F_1$, respectively. In what follows we neglect this small anisotropic splitting and use the notations $|F,F_z\rangle$ to denote the exciton states, with $F_z=-F, -F +1, \ldots F$ being the component of the total angular momentum.

The states with $F=1$ are bright, while the five states with $F=2$ are inactive in the dipole approximation, these dark states lie energetically below the bright triplet. For an exciton at motion, where its center of mass wavevector $K\ne 0$, the bright and dark states become mixed and the dark states become allowed in the dipole transitions. Disregarding cubic anisotropy effects for $\bm K\parallel z$ only states with the same component $F_z$ are mixed, making the states $|2, \pm 1\rangle$ optically active. Introducing, in analogy with $C_\alpha$ the amplitudes $D_\alpha$ $(\alpha=x,y)$ of the corresponding linear combinations of $|2, \pm 1\rangle$ states, we have, instead of the material equation~\eqref{C:eq}, the set of equations
\begin{subequations}
\label{exciton:complex:Q2}
\begin{multline}
\left( \mathcal E_{\rm exc} + \frac{\hbar^2K_z^2}{2M_{1}}- \hbar\Omega -\mathrm i \hbar\gamma \right) C_{\alpha} + \beta K_z^2 D_{\alpha} \\
=  d  E_{\alpha,\Omega}^+ + M^{(3)}_{\rm exc} E_{\alpha,\omega}^3,
\label{Calpha:Q2}
\end{multline}
\begin{multline}
\left( \mathcal E_{\rm exc} - \Delta +\frac{\hbar^2K_z^2}{2M_{2}}- \hbar\Omega  -\mathrm i \hbar\gamma \right) D_{\alpha} +\beta K_z^2 C_{\alpha}  =0,\label{Dalpha:Q2}
\end{multline}
\end{subequations}
where $\Delta$ is the dark-bright splitting at $K=0$. The effective masses $M_{1,2}$ and the mixing parameter $\beta$ can be calculated by the method developed in Ref.~\onlinecite{PhysRevB.11.3850}.  These quantities are related with the complex valence band structure of GaAs-type semiconductors.

\begin{figure}[t]
\centering
\includegraphics[width=0.45\textwidth]{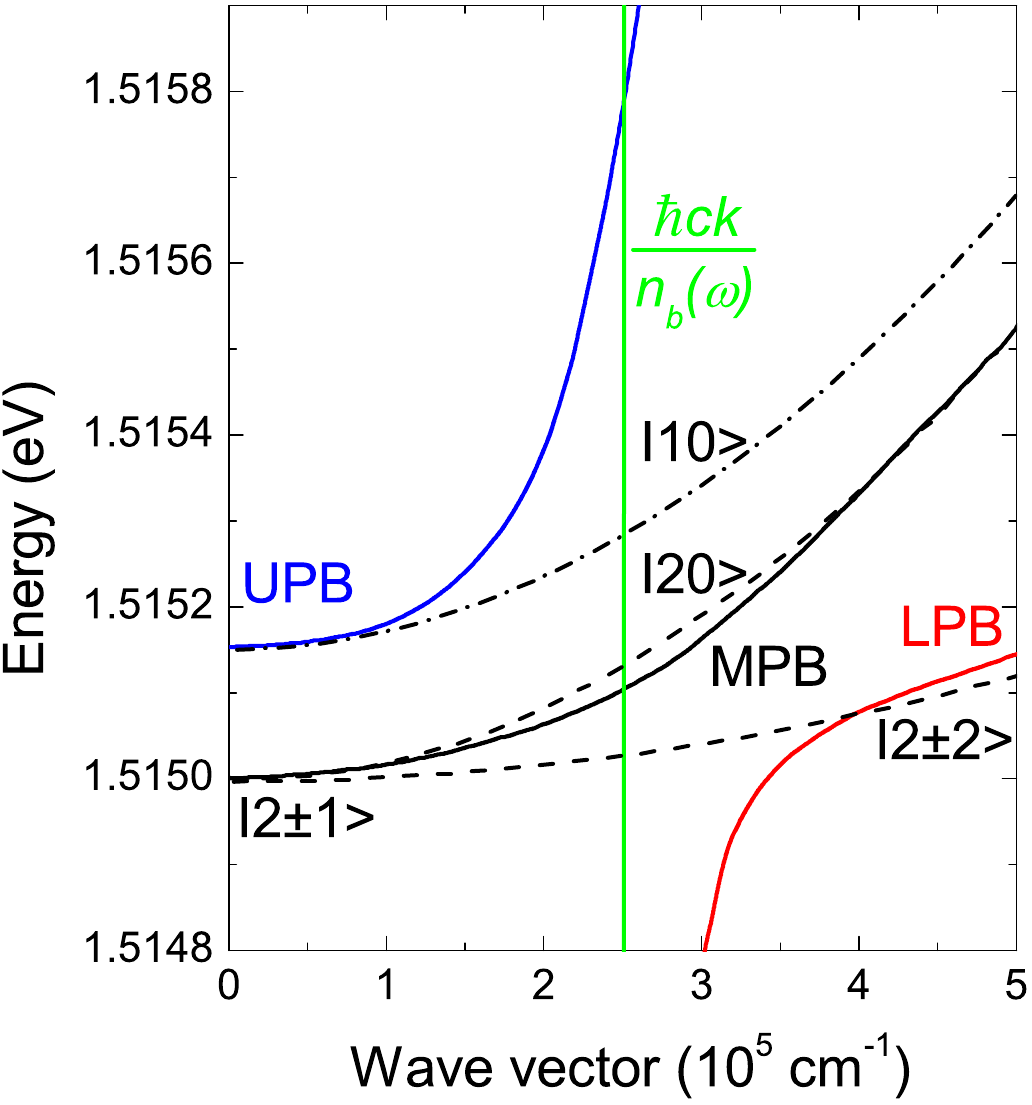}
\caption{Dispersion of exciton and exciton-polariton states in GaAs for $\mathbf{K} \parallel [001]$ after Ref.~\onlinecite{Fishman}.
Exciton-polariton dispersion for UPB, MPB and LPB is shown by solid lines. Dispersion of the longitudinal $|1,0\rangle$ exciton is given by dashed-dotted line and of the dark $|2,0\rangle$ and  $|2,\pm 2\rangle$ excitons by dashed lines.  The green line represents the light dispersion with $\hbar ck^{\omega}/n_b(\omega)$. In its crossing points with the exciton and exciton-polariton states the THG signals are principally possible, while at $B=0$ the are forbidden for the longitudinal and dark excitons with projections $0$ and $\pm 2$ and only weakly allowed for the MPB.}
\label{fig:Fig-14}
\end{figure}

Figure~\ref{fig:Fig-14} shows dispersion of exciton and exciton-polariton states in GaAs for the spectral range of $1s$-exciton. Results are plotted for $\mathbf{K} \parallel [001]$ after Figs.~1 and 3 in Ref.~\onlinecite{Fishman}, where the following parameters were chosen: energy of  $1s$-exciton without coupling with light of 1.5150~eV, dielectric constant $\epsilon=12.55$, $M=0.183$, $\hbar\omega_{LT}=80~\mu$eV~\cite{Urbrich1977}, and exchange splitting of 100~$\mu$eV. The upper, middle and lower exciton-polariton branches (UPB, MPB and LPB, respectively) are shown by solid lines. Other exciton states are not coupled with the light. The  dispersion of the longitudinal $|1,0\rangle$ exciton is traced by a dashed-dotted line and of the dark, $|2,0\rangle$ and  $|2,\pm 2\rangle$,  excitons by dashed lines.

THG signals are principally possible at the energies where the light dispersion with $\hbar ck^{\omega}/n_b(\omega)$ shown by a green line in Figure~\ref{fig:Fig-14} crosses the exciton-polariton or exciton states. 
The state $|2, \pm 1\rangle$ is dark at $K=0$.  However, at nonzero $K$, the $K^2_z$  terms induce an admixture of the $|1, \pm 1\rangle$ bright states to the $|2, \pm 1\rangle$ states. This allows their interaction with light resulting in the MPB polaritons.
Our estimates show that for the particular case of GaAs the mixing described by Eqs.~\eqref{exciton:complex:Q2} is weak and the polariton effect on the $|2,\pm 1\rangle$ states {in zero magnetic field}  is negligible. In the $T_d$ point symmetry group an additional small $\bm K$-linear mixing of the bright and dark excitons with angular momentum components $\pm 1$ is possible with the extra terms $\propto \mathrm i[\bm K\times \bm D]$ and $\mathrm i[\bm K\times \bm C]$ in Eq.~\eqref{Calpha:Q2} and Eq.~\eqref{Dalpha:Q2}, respectively. This mixing is also small and can be disregarded. The states with momentum components $0$ and $\pm 2$ are strictly forbidden at zero magnetic field. Thus, the light coupling to the $F=2$ excitons due to the complex valence band structure effects is of minor importance and the intersection points of the photon dispersion with the curves that stem from $F=2$ excitons, i.e. from intersections of light dispersion with the MPB, do not provide a sizable contribution to the THG in GaAs, see also discussion below.

\subsection{Effects of external magnetic field}
\label{subsec:magn}

Now we turn to the effects of the external magnetic field on the THG from exciton-polaritons. The magnetic field modifies the exciton states in semiconductors by inducing diamagnetic shifts of the exciton states, changing the exciton oscillator strengths, providing Zeeman splittings of excitons and also mixing dark and bright excitons. It is instructive to distinguish between the effects that are even and odd in the magnetic field $\bm B$ and also separately analyze the Faraday ($\bm B \parallel \bm K$) and the Voigt ($\bm B \perp \bm K$) geometries.

\subsubsection{Faraday geometry}\label{subsec:magn:F}

As above we assume that $\bm K\parallel z$ and $\bm B \parallel z$. For the analysis of the \emph{odd} in the field effects it is sufficient to restrict ourselves to $B$-linear contributions. The main effect here is the Zeeman effect which results in the mixing of $x$- and $y$-polarized excitons and, eventually, in the splitting of the bright exciton states into the $\sigma^+$ and $\sigma^-$ polarized states observed in the reflectivity spectra, see Fig.~\ref{fig:Fig-2}(a). Microscopically, the Zeeman effect can be described by the terms $\propto [\bm C \times \bm B]_\alpha$ in the material equation~\eqref{C:eq} for $C_\alpha$ which leads to a magnetic field-induced correction to the third-order nonlinear susceptibility (Faraday effect) as
\begin{equation}\label{eq:Faraday}
\delta {\bm P}^{3\omega} \propto {\bm P}^{3\omega}  \times {\bm B} \, ,
\end{equation}
where ${\bm P}^{3\omega}$ is given by Eq. (\ref{phenom:simple}).

Moreover, the $B$-linear term mixes the $|2,\pm 1\rangle$ states with the $|1,\pm 1\rangle$ states and additionally activates the MPB polaritons. A sufficiently strong magnetic field gives rise to two exciton branches related with the heavy-hole and light-hole states (with the momentum projection $\pm 3/2$ and $\pm 1/2$ onto the field direction, respectively). The field-activated excitons manifest themselves as additional peaks in the reflectivity spectrum at the low-energy side of the exciton resonance, see Fig.~\ref{fig:Fig-2}(a). However, for the chosen experimental conditions the contribution of these states to the THG effect is minor and unobservable in experiment. Note, that while the mixing term is linear in  $B$, the effect is \emph{even} in the magnetic field due to the zero field splitting of the exciton levels.

Now we turn to the effects \emph{even} in the magnetic field.
%
The most pronounced effects are related with the field induced variation of the exciton energy, i.e., with the diamagnetic shift, and the enhancement of the exciton oscillator strength. The diamagnetic shift can be included as the $B_z^2$-dependence of the bright exciton energy~\cite{Altarelli1973}
\begin{subequations}
\label{quadratic:B:Faraday}
\begin{equation}
\label{diamag}
\mathcal E_{\rm exc} (B_z) = \mathcal E_{\rm exc} (0)+ \alpha B_z^2,
\end{equation}
where $\alpha \sim e^2 a_B^2/\mu c^2$ with $\mu$ being the reduced electron-hole mass and $a_B$ the exciton Bohr radius. The diamagnetic shift related to the exciton Rydberg $E_B$ is controlled by the dimensionless parameter
\begin{equation}
\label{param}
\frac{\mathcal E_{\rm exc} (B_z) - \mathcal E_{\rm exc} (0)}{E_B} \sim \frac{a_B^4}{l_B^4},
\end{equation}
here $l_B = \sqrt{\hbar c/e|B_z|}$ is the magnetic length. Hence, the larger the exciton Bohr radius, the stronger is the effect of magnetic field. It is also important that the magnetic field significantly modifies the $1s$ exciton wavefunction by shrinking it in the $(xy)$ plane. As a result, the exciton wavefunction at the coinciding electron and hole coordinates $\Phi_{1s}(0)$ increases resulting in the increase of the exciton oscillator strength,
\begin{equation}
\label{osc}
\hbar\omega_{LT}(B_z) = \hbar\omega_{LT} (0)+ \alpha' B_z^2,
\end{equation}
with parameter $\alpha'$. The relative variation of the exciton oscillator strength
\begin{equation}
\label{param1}
\frac{\hbar\omega_{LT}(B_z) - \hbar\omega_{LT} (0)}{ \hbar\omega_{LT} (0)} = \frac{|\Phi_{1s}(B_z)|^2 - |\Phi_{1s}(0)|^2}{|\Phi_{1s}(0)|^2} \sim \frac{a_B^4}{l_B^4},
\end{equation}
\end{subequations}
is controlled by the same dimensionless parameter as  the diamagnetic energy shift, Eq.~\eqref{param}. We note that in GaAs at $B=10$~T the ratio $a_B/l_B \approx 3$ and the perturbation approach in Eqs.~\eqref{diamag} and \eqref{osc} is inapplicable. Correspondingly, the strong diamagnetic shift (of about 6~meV) and a substantial increase of the $\hbar\omega_{LT}$ by a factor of $\sim 6$ in GaAs are clearly visible in the reflectivity data, Fig.~\ref{fig:Fig-2}(c). For CdTe and ZnSe the effect is accordingly smaller due the smaller Bohr radii in these semiconductors, see insets in Figs.~\ref{fig:Fig-11} and \ref{fig:Fig-12}.

The rotational anisotropy of the third harmonic generation in the Faraday configuration is simple. Within our model the induced polarization $\bm P^{3\omega}$ follows the incident field $\bm E^\omega$ but, due to the Faraday effect described by Eq.~\eqref{eq:Faraday}, $\bm P^{3\omega}$ is generally no longer parallel to $\bm E^\omega$. Thus for the configuration with parallel polarizer and analyzer ($\bm E^{3\omega}\parallel \bm E^\omega$) the THG intensity is independent of their common orientation. A signal in the crossed polarizations $\bm E^{3\omega} \perp \bm E^\omega$ is expected due to the Faraday effect, it vanishes at $B=0$ and is also invariant under the polarizer/analyzer rotation.

\subsubsection{Voigt geometry} \label{subsec:magn:V}

A transversal magnetic field ($\bm B \perp \bm K$, $\bm B \parallel x$) can also activate longitudinal and dark excitons with projections $0$ and  $\pm 2$ on the direction of $\mathbf{K}$. For instance, in the $B$-linear regime, due to the Zeeman effect, the longitudinal exciton with $C_z \ne 0$ is mixed with the transverse excitons acquiring a non-zero oscillator strength. In this way, the exciton $|1,0\rangle$ became active both in the linear optical response and in the THG. To describe this mixing it is sufficient to add in Eq.~\eqref{C:eq} terms of the form $\propto [\bm C \times \bm B]_\alpha$, which at $\bm B \parallel x$ mix the $C_y$ and $C_z$ amplitudes (similarly as $C_x$ and $C_y$ for $\bm B \parallel z$). The contribution of the $|1,0\rangle$  state is expected to be polarized perpendicular to the magnetic field, $\bm P^{3\omega} \perp \bm B$.

Furthermore, the magnetic field in the Voigt geometry also mixes the $F=1$ and $F=2$ excitons. The contribution of the $|2,0\rangle$ state mixed with the bright state is expected to be polarized parallel to the magnetic field, $\bm P^{3\omega} \parallel \bm B$. The dark $|2,\pm 2 \rangle$ states are mixed with the bright  $|1,\pm 1 \rangle$ ones acquiring thereby the polarization properties of the latter. Again, although mixing is induced by the $B$-linear terms, the effect is \emph{even} due to the zero field fine energy splitting.

A further increase of the magnetic field gives rise to further mixing and splitting of the states as well as, similarly to the Faraday geometry, to the diamagnetic shift, Eq.~\eqref{diamag}, and the variation of the oscillator strength, Eq.~\eqref{osc}, \emph{even}, with magnetic field. Here we present only the results for strong enough magnetic fields, where the diamagnetic shifts of the exciton states exceed by far $\hbar\omega_{LT}$. In this regime one can disregard, in first approximation, the fine structure of the excitons at $B=0$ as well as the exciton dispersion (i.e. set $K=0$) and use the approach of Ref.~\onlinecite{Altarelli1973}. Thus, the octuplet of exciton states gives rise to two states with the microscopic dipole moment $\bm P^{3\omega} \parallel \bm B$ and four states with $\bm P^{3\omega} \perp \bm B$. We neglect for simplicity the splitting within the doublet and quartet of states and focus only on quadratic in magnetic field effects\cite{Altarelli1973}. The analysis shows that, for the state polarized perpendicular to the magnetic field, the dependence of the diamagnetic shift and the oscillator strength on the field is the same as for the Faraday geometry, Eqs.~\eqref{quadratic:B:Faraday}. By contrast, for the state polarized parallel to the field, the diamagnetic shift and the variation of oscillator strength are different. As a result, instead of Eqs.~\eqref{quadratic:B:Faraday} for $\bm B\parallel x\parallel [100]$ we have
\begin{subequations}
\label{quadratic:B:Voigt}
\begin{equation}
\label{diamag:V}
\mathcal E_{\rm exc} (B_x) = \mathcal E_{\rm exc} (0)+
\begin{cases}
 \alpha B_x^2,\quad \bm P^{3\omega} \parallel y,\\
 \alpha_\parallel B_x^2,\quad \bm P^{3\omega} \parallel x,\\
\end{cases}
\end{equation}
\begin{equation}
\label{osc:V}
\hbar\omega_{LT}(B_x) = \hbar\omega_{LT} (0)+
\begin{cases}
 \alpha' B_x^2,\quad \bm P^{3\omega} \parallel y,\\
 \alpha_\parallel' B_x^2,\quad \bm P^{3\omega} \parallel x,\\
\end{cases}
\end{equation}
\end{subequations}
with the additional parameters $\alpha_\parallel$ and $\alpha_\parallel'$. The calculations\cite{Altarelli1973} show that $\alpha>\alpha_\parallel$. As a result, the higher state in the Voigt geometry is polarized perpendicular to the magnetic field and diamagnetic shift is the same as for the Faraday geometry. The lower state is polarized parallel to the magnetic field.
This is in agreement with our experimental results in Figs.~\ref{fig:Fig-4}, \ref{fig:Fig-7}(e) and \ref{fig:Fig-7}(h). This polarization dependence is also apparent in the experimental data on reflectivity, see Fig.~\ref{fig:Fig-2}.

Due to the fact that, in the considered model, the states are active either for $\bm P^{3\omega} \perp \bm B$ or for $\bm P^{3\omega} \parallel \bm B$, the THG rotational anisotropies have simple forms. For instance, for $\bm E^{3\omega}\parallel \bm E^{\omega}$
\begin{subequations}
\label{rot:Voigt}
\begin{equation}
\label{co}
I^{3\omega}_\parallel (\varphi) \propto
\begin{cases}
\cos^4{\varphi},\quad \bm P^{3\omega} \parallel y,\\
\sin^4{\varphi}, \quad \bm P^{3\omega} \parallel x,\\
\end{cases}
\end{equation}
where $\varphi$ is defined in Fig.~\ref{fig:Fig-1}, while for $\bm E^{3\omega}\perp \bm E^{\omega}$ four-fold symmetry
\begin{equation}
\label{cross}
I^{3\omega}_\perp \propto \cos^2{\varphi}\sin^2{\varphi} =\frac{1}{8}(1-\cos{4\varphi}),
\end{equation}
\end{subequations}
is expected. The same rotation anisotropy is expected for the THG signal if it comes from the $|1,0\rangle$ and $|2,0\rangle$ activated states.

Let us summarize the results of the magnetic field effects on the THG involving exciton-polariton states. The THG intensity increase is expected to be provided by the growth of the exciton oscillator strength. The spectral shift of THG line to higher energies is mainly contributed by the exciton diamagnetic shift, but also partly by repulsion of the UPB and LPB polaritons due to increase of the oscillator strength. The stronger changes in magnetic field are expected for materials having excitons with larger Bohr radius and smaller exciton binding energy. The obvious scaling factor here is the ratio of the exciton Bohr radius to the magnetic length.

\section{Discussion}
\label{sec:discussion}

In this section we present the comparison between the experimental data and the developed theory and discuss the obtained results. First, we focus on the main effects which largely do not depend on the experimental geometry: The THG intensity strongly increases with growing magnetic field, Figs.~\ref{fig:Fig-4} and \ref{fig:Fig-5}. The largest effect is observed in GaAs and becomes progressively weaker in CdTe and ZnSe, where the exciton Bohr radii are smaller. A qualitative explanation already follows from Eqs.~\eqref{osc} and \eqref{osc:V}: The magnetic field shrinks the exciton wavefunction and increases its oscillator strength. While the dip in the reflectivity spectrum $\propto |\Phi_{1s}(0)|^2$ the intensity of the third harmonic is proportional to the $|\Phi_{1s}(0)|^4$, see Eqs.~\eqref{chi3:exc}, \eqref{THG:resonance} and \eqref{no:space:1}. Thus, the approximately $6$-fold increase of the amplitude of the exciton resonance in the reflectivity, Fig.~\ref{fig:Fig-2}(b), should result in an about $36$-fold increase of the THG intensity. This is in good agreement with the experimental data in  Figs.~\ref{fig:Fig-4} and \ref{fig:Fig-5}, where enhancement ranging from 25 to 50 times is seen. In accordance with Eqs.~\eqref{param} and \eqref{param1} the enhancement of the exciton oscillator strength and, therefore, the THG intensity is controlled by the ratio of the Bohr radius to the magnetic length. Correspondingly, in CdTe and ZnSe, where the excitons Bohr radii are smaller, the THG enhancement is weaker: It is about $3$-fold in CdTe, Fig.~\ref{fig:Fig-11}, and is absent in ZnSe, Fig.~\ref{fig:Fig-12}.

In order to provide a quantitative comparison with experiment in GaAs, we have extracted the relative values of the exciton oscillator strength and diamagnetic shift as a function of magnetic field from the amplitude of the exciton resonance in the reflectivity, Fig.~\ref{fig:Fig-2}(b). Then, we calculated after Eq.~\eqref{THG:resonance} the THG spectra and their evolution in magnetic field. From these spectra we extracted the magnetic field dependencies of the THG enhancement factor and the THG peak energy. They are given by the dashed-dotted lines in Figs.~\ref{fig:Fig-5} and \ref{fig:Fig-6} for comparison with experiment. The agreement between experiment and theory is good. Note, that for a fully quantitative description of the data, i.e., for accounting for the difference between the Faraday and Voigt geometries, the exciton fine structure needs to be fully taken into consideration, which is beyond the scope of the present paper.

The THG rotational anisotropies shown Fig.~\ref{fig:Fig-7} are in line with our model predictions. At zero magnetic field, in agreement with Eqs.~\eqref{rotation} and \eqref{chi3:exc}, the THG signal is detected only in the parallel configuration of polarizer and analyzer ($\bm E^{3\omega}\parallel \bm E^\omega$) and is absent in the perpendicular configuration ($\bm E^{3\omega}\perp \bm E^\omega$), see Figs.~\ref{fig:Fig-7}(a) and \ref{fig:Fig-7}(b). For the parallel configuration an isotropic diagram with $\mathcal A =\mathcal C$ is expected. The experimental data in Fig.~\ref{fig:Fig-7}(a) slightly deviate from an isotropic dependence giving the ratio $\mathcal C/\mathcal A \approx 0.82$. The deviation might be due to the contribution of the remote bands, which are neglected in our model.

In the Faraday geometry the strongly enhanced THG signal in the parallel configuration becomes perfectly isotropic, Fig.~\ref{fig:Fig-7}(c). It is well fitted with $\mathcal A = \mathcal C$. Additionally, a small signal appears in the $\bm E^{3\omega} \perp \bm E^\omega$ configuration, Fig.~\ref{fig:Fig-7}(d), due to the Faraday effect, i.e., the rotation of the THG polarization plane described by Eq.~\eqref{eq:Faraday}.

The anisotropies of the magnetic-field-enhanced THG signals in the Voigt geometry are also in good agreement with the developed model. More specifically, in agreement with Eqs.~\eqref{quadratic:B:Voigt} and \eqref{rot:Voigt} the two-fold rotational symmetry in the parallel configuration, $\bm E^{3\omega} \parallel \bm E^{\omega}$, is observed. For the higher in energy exciton state the intensity is maximal for the polarization perpendicular to $\bm B$ [Fig.~\ref{fig:Fig-7}(e)], while for the lower one it is maximal for the polarization parallel to $\bm B$ [Fig.~\ref{fig:Fig-7}(h)]. In the $\bm E^{3\omega}\perp \bm E^\omega$ (perpendicular) configuration the rotational anisotropy for both exciton states has the same four-fold shape,  Figs.~\ref{fig:Fig-7}(f) and \ref{fig:Fig-7}(g). This is in agreement with the model predictions of Eq.~\eqref{cross}.

For understanding all the details of the resonant third harmonic 
generation in bulk semiconductors the exciton-polariton effect plays an 
important role. Particularly, the energy of the THG resonance in the 
spectrum is determined mainly by the polariton effects, see Eq.~\eqref{Omega*} and Fig.~\ref{fig:Fig-13}, and particularly sensitive to the dispersion of the background 
dielectric constant. However, the basic features of the THG intensity 
and polarization dependence on magnetic field can be understood in a 
simplified approach, where the polariton effects are disregarded. 
Particularly, as demonstrated above, the magnetic field enhancement of 
the THG and the rotational anisotropies of the THG are readily 
understood by accounting for the exciton contribution only. Finally, we 
note that the strength of exciton-polariton effect is controlled by the 
ratio of the exciton oscillator strength, $\omega_{LT}$, and the exciton 
damping, $\gamma$. That is why the exciton-polariton effect in the 
studied semiconductors is important for the optically active $1s$ ground 
exciton, which has the largest oscillator strength.

\section{Conclusions}

In conclusion, magnetic-field-induced THG with well-defined polarization properties and characteristic magnetic-field
dependencies is observed on the exciton-polariton resonance in bulk GaAs. Although the THG is allowed in the electric-dipole approximation, a strong enhancement of the THG intensity by a factor of fifty at 10~T is found at the $1s$-exciton resonance. The effect is much weaker in CdTe and ZnSe having the same crystal symmetry, but larger exciton binding energies.
We developed a microscopic theory of THG on exciton-polaritons, which accounts for the resonant properties, and the modification of the exciton-polariton parameters in external magnetic field. This theory shows that the resonant enhancement of THG in the vicinity of the $1s$-exciton resonance is provided by the fulfillment of the THG phase matching condition due to the exciton-polariton dispersion. The magnetic field enhancement of the THG intensity is directly related to the increase of the exciton oscillator strength, which is large in materials with smaller exciton binding energy and, therefore, larger Bohr radius.

\textbf{Acknowledgements.} The authors are thankful to D.~Fr\"ohlich for stimulating discussions. This work was supported by the Deutsche Forschungsgemeinschaft via ICRC TRR160 (project C8) and SFB TRR142 (project B01), and the Russian Foundation for Basic Research (Project No. 15-52-12015).

\end{document}